\documentstyle[11pt,epsf]{article}

\newcommand{\newsection}{ \setcounter{equation}{0} \section}

\newcommand{\beq}{\begin{equation}} \newcommand{\eeq}{\end{equation}}
\newcommand{\bea}{\begin{eqnarray}} \newcommand{\eea}{\end{eqnarray}}

  \newcommand
{\Romannumeral}[1]{\uppercase\expandafter{\romannumeral#1}}

\newcommand{\be}{\begin{enumerate}} \newcommand{\ee}{\end{enumerate}}
\newcommand{\bi}{\begin{itemize}} \newcommand{\ei}{\end{itemize}}
\newcommand{\ba}{\begin{array}} \newcommand{\ea}{\end{array}}
\newcommand{\bc}{\begin{center}} \newcommand{\ec}{\end{center}}
\newcommand{\bt}{\begin{tabular}} \newcommand{\et}{\end{tabular}}

\def\lsim{\mathrel{\rlap{\lower4pt\hbox{\hskip1pt$\sim$}}
    \raise1pt\hbox{$<$}}}           
\def\gsim{\mathrel{\rlap{\lower4pt\hbox{\hskip1pt$\sim$}}
    \raise1pt\hbox{$>$}}}           

\newcommand{\half}{\textstyle {1\over2} \displaystyle}    
\newcommand{\third}{\textstyle {1\over3} \displaystyle}   
\newcommand{\quarter}{\textstyle {1\over4} \displaystyle} 
\newcommand{\sixth}{\textstyle {1\over6} \displaystyle}   
\newcommand{\Dslash}{{\hbox{D}\kern-0.6em\raise0.15ex\hbox{/}}} 

 \newcommand{\k}{\kappa}

\begin{document}

\setlength{\oddsidemargin}{0cm} \setlength{\baselineskip}{7mm}

\input epsf

\begin{normalsize}\begin{flushright}
DAMTP-2005-59 \\
June 2005 \\
\end{flushright}\end{normalsize}

\begin{center}
  
\vspace{5pt}
  


{\Large \bf Nonlocal Effective Gravitational Field Equations}

{\Large \bf and the Running of Newton's G }

\vspace{10pt}

\vspace{10pt}

{\sl H. W. Hamber}
$^{}$\footnote{e-mail address : HHamber@uci.edu} \\
Department of Physics and Astronomy \\
University of California \\
Irvine, CA 92697-4575, USA \\

\vspace{5pt}

and

\vspace{5pt}

{\sl R. M. Williams}
$^{}$\footnote{e-mail address : R.M.Williams@damtp.cam.ac.uk} \\
Girton College, Cambridge CB3 0JG, and   \\
Department of Applied Mathematics and Theoretical Physics \\
Centre for Mathematical Sciences \\
Wilberforce Road, Cambridge CB3 0WA, United Kingdom.
\\

\end{center}

\begin{center} {\bf ABSTRACT } \end{center}

\noindent

Non-perturbative studies of quantum gravity have recently 
suggested the possibility that the strength of gravitational
interactions might slowly increase with distance.
Here a set of generally covariant effective field equations are proposed, which
are intended to incorporate the gravitational, 
vacuum-polarization induced, running of Newton's constant $G$.
One attractive feature of this approach is that,
from an underlying quantum gravity perspective, the resulting long distance
(or large time) effective gravitational action inherits only one
adjustable parameter $\xi$, having the units of a length, 
arising from dimensional transmutation in the gravitational sector.
Assuming the above scenario to be correct, some simple predictions for
the long distance corrections to the classical standard model
Robertson-Walker metric are worked out
in detail, with the results formulated as much as possible in a 
model-independent framework.
It is found that the theory, even in the limit of vanishing renormalized
cosmological constant, generally predicts an accelerated
power-law expansion at later times $t \sim \xi \sim 1/H$.




\vfill

\pagestyle{empty}

\newpage

\pagestyle{plain}

\vskip 10pt
\newsection{Introduction}
\hspace*{\parindent}

Non-perturbative studies of quantum gravity have recently suggested
the possibility that gravitational couplings might be weakly scale
dependent due to nontrivial renormalization group effects.
This would introduce a new gravitational scale, unrelated to Newton's
constant, required in order to parametrize the gravitational running
in the infrared region. 
If one is willing to accept such a scenario, then it seems
difficult to find a compelling theoretical
argument for why the non-perturbative scale entering the coupling
evolution equations should be very small, comparable to the Planck length.
One possibility put forward recently is that
the relevant non-perturbative scale is related to the curvature
and therefore macroscopic in size, which could
have observable consequences. 
One key ingredient in this argument is the relationship, to some extent
supported by 
Euclidean lattice results combined with renormalization group arguments,
between the scaling
violation parameter and the scale of the average curvature. 
Irrespective of the specific details of a gravitational theory
at very short distances, such results would bring gravitation 
more in line with the rest of the Standard Model, where all gauge
couplings are in fact known to run.

In this paper we investigate the effects of a running gravitational
coupling $G$ at large distances, with as few assumptions as possible
about the ultimate behavior of the theory at extremely short distances,
where several possible scenarios include a string
cutoff at length scales $\lambda_S = (2 \pi \alpha')^{1/2}$ \cite{scosm},
the appearance of higher derivative terms (either
as direct contributions or as radiative corrections), or
perhaps a - somewhat less appealing - explicit ultraviolet
cutoff at the Planck scale.
The running of the gravitational coupling will generally be
assumed to be driven by graviton
vacuum polarization effects, which produce an anti-screening effect
some distance away from the primary source, and therefore tend
to increase the strength of the gravitational coupling.
The above scenario is quite different from what one would expect for example
in supergravity theories, where significant cancellations
arise in perturbation theory between graviton and matter loops \cite{sugra},
and in contrast to ordinary gravity where in weak field perturbation
theory $L$ loops contribute $L+1$
powers of the curvature tensor to the effective action \cite{thooft}.
Instead, the running of Newton's constant is thought to arise due
to the presence of a non-trivial, genuinely non-perturbative,
ultraviolet fixed point \cite{wilson,parisi,sym} 
(a phase transition in statistical mechanics parlance \cite{wilson}).

In this paper a power law (as opposed to a logarithmic) running of $G$ will
be implemented via manifestly covariant nonlocal terms in the effective
gravitational action and field equations.
It ultimately will involve the inverse of the covariant d'Alembertian
raised to some fractional power $1/ 2\nu$, which in the framework of
the present paper remains largely unspecified, although non-perturbative 
models for quantum gravity have recently put forward some rather
specific predictions. 

Let us recall here, to provide some degree of motivation, the recent
discussions of ~\cite{det,critical} as a possible
theoretical framework for the running of Newton's $G$.
The above results suggest that the gravitational
constant $G$ cannot be regarded a constant as in the classical theory,
but instead changes slowly with scale due to the presence
of weak gravitational vacuum polarization effects, in a way described by
\beq
G(r) \; = \; G(0) \left [ \; 1 \, + \, c_\xi \, ( r / \xi )^{1 / \nu} \, 
+ \, O (( r / \xi )^{2 / \nu} ) \; \right ]
\label{eq:grun}
\eeq
The exponent $\nu$, generally related to the derivative
of the beta function for pure gravity evaluated at the non-trivial ultraviolet
fixed point via the relation $ \beta ' (G_c) \, = \, - 1/ \nu $,
is supposed to universally characterize the long-distance properties of quantum
gravitation, and is therefore expected to be independent of the specifics
related to the nature of the ultraviolet regulator, or other detailed
short distance features of the theory.
\footnote{Already in ordinary Einstein gravity one finds for very short distances
$r\sim l_P$ corrections to the static potential, which can be computed
perturbatively ~\cite{donoghue}. In general for such short
distances string corrections and/or higher
derivative terms should be considered as well.}

Recent estimates for the value of the universal scaling dimension
$\nu^{-1} = - \beta ' (G_c)$
derived from non-perturbative studies of gravity vary from 
$\nu^{-1} \approx 3.0$ ~\cite{critical} in the Euclidean Regge lattice case,
to $\nu^{-1} \approx 3.8$ in the $2+\epsilon$ expansion
\cite{sigma} about two dimensions carried 
to two loops ~\cite{epsilon,epsilon1,epsilon2,epsilon3}, 
to $\nu^{-1} \approx 2.7$ ~\cite{litim} and 
$\nu^{-1} \approx 1.7$ ~\cite{reuter} in an approximate
renormalization group treatment a la Wilson based on an Einstein-Hilbert
truncation,
with some significant uncertainties in all three approaches.
More details, as well as a systematic comparison of the various methods
and estimates, can be found in ~\cite{ttmodes}, where we argued, based
on geometric arguments, in
favor of the exact value of the exponent $\nu = 1/3$ for pure gravity 
in four dimensions, and $O(1/(d-1))$ for large $d$.
It is perhaps a testament to how far these calculations have progressed that
actual numbers have emerged which can meaningfully be compared between
different (lattice and continuum) approaches.
It should also be noted that, from a quantum gravity perspective, there
are really no adjustable parameters in Eq.~(\ref{eq:grun}), 
except for the new
non-perturbative curvature scale $\xi$:
both $c_{\xi}$ and $\nu$ are in principle finite and calculable numbers.

The mass scale $m = \xi^{-1}$ in Eq.~(\ref{eq:grun}) is supposed to
determine the
magnitude of quantum deviations from the classical theory, and separates
the short distance,
ultraviolet regime with characteristic momentum scale $ \mu \ll m $, where
non-perturbative quantum corrections are negligible, from
the long distance regime where quantum corrections become significant.
The magnitude of $\xi$ itself
involves, in a rather non-trivial way, the dimensionless
bare coupling $G$, the fixed point value
$G_c$ and the ultraviolet cutoff $\Lambda$,
\beq
\xi^{-1} \, \propto \, \Lambda \,
\exp \left ( { - \int^G \, {d G' \over \beta (G') } } \right )
\, \mathrel{\mathop\sim_{G \rightarrow G_c }} \,
\Lambda \, | \, G - G_c |^{ - 1 / \beta ' (G_c) } \;\;\; .
\label{eq:xi}
\eeq
Ultimately to make progress and determine the actual physical value for the
non-perturbative
scale $\xi$ some physical input is needed, as the underlying theory
{\it cannot} fix it (the ratio of the physical Newton's constant
to $\xi^2$ can be as small as one desires, provided the bare coupling $G$
is very close to its fixed point value $G_c$).
It seems natural to identify  $ 1 / \xi^2 $ with either some 
very large average spatial curvature scale,
or perhaps more appropriately with the Hubble 
constant (as measured today) determining the macroscopic expansion rate of the
universe via the correspondence
\beq
\xi \; = \; 1 / H \;\; ,
\label{eq:hub}
\eeq
in a system of units for which the speed of light equals one.
\footnote{
A possible scenario is one in which 
$\xi^{-1} = H_\infty = \lim_{t \rightarrow \infty} H(t) =
\sqrt{\Omega_\Lambda}\, H_0$
with $H_\infty^2 = { 8 \pi G \over 3 } \lambda = { \Lambda \over 3 } $,
where $\lambda$ is the observed cosmological constant, and for which
the horizon radius is $R_{\infty} = H_{\infty}^{-1}$.}

Let us briefly digress here, and
recall that in non-Abelian $SU(N)$ gauge theories a similar set
of results is known to hold for the renormalization-group induced
running of the gauge coupling $g$,
so it will be instructive to draw further on the analogy with
QCD, and non-Abelian gauge theories in general.
Of course one crucial difference between gravity and ordinary gauge theories
lies in the fact that, in the latter case, the
evolution of the coupling constant can be systematically computed
in perturbation theory due to asymptotic freedom, a 
statement which is known to reflect the fact that such theories become
non-interacting at short distances, up to logarithmic corrections, making
perturbation theory consistently applicable.
It is well known that for weak enough gauge coupling 
in $SU(N)$ gauge theories one has
\beq
{ 1 \over g^2 (\mu) } \; = \; { 1 \over g^2 ( \Lambda_{\overline{MS}} ) } 
\; + \; 2 \beta_0 \; \log \left (
{ \mu \over \Lambda_{\overline{MS}} } \right ) \; + \cdots
\label{eq:qcdrun}
\eeq
with $\beta_0$ the coefficient of the lowest order term in the
beta function, $\mu=1/r$ an arbitrary momentum scale,
$\Lambda_{\overline{MS}} \approx 220 MeV$ a non-perturbative scale parameter,
and the dots denoting higher loop effects.
Instead of the $\Lambda_{\overline{MS}}$ parameter
one could just as well use some other physical scale, such
as the inverse of the gauge correlation length, $m_{0^{++}} = \xi^{-1}$, where
the $0^{++}$ denotes the lowest glueball state (the Slavnov-Taylor
identities prevent of course the gluon from acquiring a mass to any order
in perturbation theory).
For the purpose of comparing to gravity,
one should perhaps emphasize that confining non-abelian gauge
theories such as QCD do not, and cannot,
directly determine the scale
$\Lambda_{\overline{MS}}$, 
which needs to be ultimately fixed by experiment from
say a direct measurement of the size of scaling violations.
Its magnitude involves in a non-trivial way the bare gauge coupling $g$ and
the ultraviolet cutoff $\Lambda$,
\beq
\Lambda_{\overline{MS}} \; \propto \; \Lambda \, 
\exp \left ( { - \int^g \, {d g' \over \beta (g') } } \right )
\eeq
which is very much analogous to Eq.~(\ref{eq:grun}).
The correspondence with QCD and nonabelian gauge theories
would therefore suggest 
$\xi^{-1} \leftrightarrow \Lambda_{\overline{MS}}$, with the
gravitational $\xi$ a new
non-perturbative scale, ultimately also to be determined from experiment.

Although not always necessarily advantageous
(most perturbative calculations,
being based on Feynman diagrams,
are eventually done in momentum space and do not seem to
benefit significantly from this approach),
the running of the gauge coupling $g$ can be re-formulated in
terms of an effective action, involving the d'Alembertian acting on
functions of the field strength.
One sets
\beq
{ 1 \over g^2 ( \Box ) } \; = \; { 1 \over g^2 ( \Lambda_{\overline{MS}} ) } 
\; + \; \beta_0 \; \log \left (
{ \Box \over \Lambda_{\overline{MS}}^2 } \right ) \; + \cdots
\eeq
with $2 \beta_0 = (11 N - 2 n_f)/ (24 \pi^2) $ for non-abelian
$SU(N)$ gauge theories with $n_f$ massless fermion flavors, and  
with the log of the d'Alembertian $\Box$ suitably defined, for example, via 
\beq
\log \left (
{ \Box \over \mu^2 } \right ) 
\; = \; \int_0^\infty d m^2 \, \left \{ 
{ 1 \over m^2 \, + \, \mu^2 }
\, - \, { 1 \over m^2 \, + \, \Box } \right \}
\eeq
leading to a one-loop corrected effective action of the form
\beq
I_{eff} \; = \; \quarter \int dx \, F_{\mu\nu} (x)
\, \left (
{ 1 \over g_0^2 } \; + \; \beta_0 \; \log \left (
{ \Box \over \mu^2 } \right ) \; + \; \cdots \right) \,
F^{\mu\nu} (x) 
\eeq
with $\mu$ an appropriately chosen mass scale ~\cite{vilkogauge}.

In the gravitational case the corrections described by
Eq.~(\ref{eq:grun}) have a more complicated structure,
and in particular are no longer logarithmic.
But they can be viewed for example as arising from a resummation of an infinite
number of loop logarithms, as in the expansion
\beq
\sum_{n=0}^{\infty} \, {\textstyle {1 \over n !} \displaystyle} 
\left ( - {\textstyle {1 \over 2 \nu } \displaystyle} \right )^n 
\left ( \log \xi^2 \Box  \right )^n
\; = \;
\left ( { 1 \over \xi^2 \Box } \right )^{ 1 / 2 \nu} 
\eeq
In the next section we shall describe how the renormalization group
induced running of the gravitational constant can be implemented
in a simple way via a non-local set of manifestly covariant correction terms
arising in the effective, long distance gravitational field equations.
These effective equations can then be used as a basis for a systematic
discussion of various quantum corrections to the standard solutions
of the classical field equations.

\newpage

\vskip 30pt
\newsection{Effective Gravitational Action and Effective Field Equations}
\hspace*{\parindent}

In general terms, a quantum-mechanical running of the gravitational
coupling implies the replacement
\beq
G \;\; \rightarrow \;\; G(r)
\eeq
in classical physical observables.
This is easier said than done, as in gravity the $r$ in the running
coupling $G(r)$ is coordinate dependent,
and as such can lead to considerable ambiguities regarding the interpretation
of exactly which distance $r$ is involved.
A more satisfactory approach would replace $G(r)$ in the gravitational action 
\beq
I \; = \; { 1 \over 16 \pi \, G } \int dx \sqrt{g} \, R
\eeq
with a manifestly covariant object, intended to correctly represent an
invariant distance, and incorporating the running of $G$ as expressed in 
Eq.~(\ref{eq:grun}),
\beq
\rightarrow \;\;\;\;
{ 1 \over 16 \pi \, G } \int dx \sqrt{g} \,
\left( 1 \, - \, c_{\Box} \, 
\left ( { 1 \over \xi^2 \Box } \right )^{ 1 / 2 \nu} \, 
+ \, O (( \xi^2 \Box )^{- 1 / \nu} ) \; \right ) R
\label{eq:ieff_r}
\eeq
with the covariant d'Alembertian operator $\Box$ defined through an
appropriate combination of covariant derivatives
\beq
\Box \; = \; g^{\mu\nu} \nabla_\mu \nabla_\nu 
\eeq
Multiplication by the coordinate $r$ gets therefore replaced
by the action of $\Box$, whose Green's function in $D$ space-time dimensions
is known to behave as
\beq
< x | \, { 1 \over \Box } \, | y > \delta ( r - d(x,y \, \vert \, g) ) 
\; \sim \; {1 \over r^{D-2} }
\eeq
Here $d$ would be the distance along a minimal path $z^{\mu}(\tau)$
connecting the points $x$ and $y$ in a fixed
background geometry characterized by the metric $g_{\mu\nu}$,
and given by
\beq
d(x,y \, \vert \, g) \; = \;
\int_{\tau(x)}^{\tau(y)} d \tau 
\sqrt{ \textstyle g_{\mu\nu} ( z )
{d z^{\mu} \over d \tau} {d z^{\nu} \over d \tau} \displaystyle } \;\; .
\label{eq:distance}
\eeq
As a result $1 / \Box $ can be envisioned as a coordinate independent
way of defining consistently what is meant by $r$ in the running of $G(r)$,
\beq
G (r) \;\; \rightarrow \;\; G( \Box )
\label{eq:gbox}
\eeq
The above prescription has in fact been used successfully for some
time to systematically incorporate the effects of radiative corrections
in an effective action formalism \cite{vilko,vilkorep,bmv}.
It should be noted that
the coefficient $c_\xi$ in Eq.~(\ref{eq:grun})
is expected to be a calculable
number of order one, but not necessarily the same
as the coefficient $c_\Box$, as $r$ and $1 / \sqrt{\Box}$ are clearly
rather different entities to begin with.
\footnote{
In the lattice theory
$c_\xi$ was originally estimated from the invariant curvature
correlations at around $c_\xi \approx 0.01$, while more
recently it was estimated at $c_\xi \approx 0.06$
from the correlation of Wilson lines ~\cite{ttmodes}.
It is important to note that while the exponent $\nu$ is universal,
$c_\xi$ in general depends on the specific choice of regularization scheme
(i.e. lattice regularization versus dimensional regularization or
momentum subtraction scheme).}

One should recall here that in general the form of the covariant
d'Alembertian operator $\Box$
depends on the specific tensor nature of the object it is acting on,
\beq
\Box \; T^{\alpha\beta \dots}_{\;\;\;\;\;\;\;\; \gamma \delta \dots}
\; = \; g^{\mu\nu} \nabla_\mu \left( \nabla_\nu \;
T^{\alpha\beta \dots}_{\;\;\;\;\;\;\;\; \gamma \delta \dots} \right)
\eeq
Thus on scalar functions one obtains the fairly simple result
\beq
\Box \, S(x) \; = \; 
{1 \over \sqrt{g} } \, \partial_\mu \, g^{\mu\nu} \sqrt{g} \, \partial_\nu
\, S(x)
\eeq
whereas on second rank tensors one has the significantly
more complicated expression
$\Box T_{\alpha\beta} \, \equiv \, 
g^{\mu\nu} \nabla_\mu (\nabla_\nu T_{\alpha\beta}) $.
Furthermore one should recognize that the form for the
effective gravitational action of Eq.~(\ref{eq:ieff_r})
is possibly not unique.
A more integration-by-parts symmetric expression would be, for example,
\beq
I \; = \; { 1 \over 16 \pi G } \int dx \sqrt{g} \; \sqrt{R} 
\left ( 1 \, - \, c_{\Box} \, 
\left ( {1 \over \xi^2 \Box } \right )^{ 1 / 2 \nu} \, 
+ \dots \right ) \sqrt{R} 
\label{eq:ieff_sr}
\eeq
In general the covariant operator appearing in the above expression, namely
\beq
A (\Box) \; = \; c_{\Box} 
\left( { 1 \over \xi^2 \Box } \right)^{1/2\nu} 
\eeq
has to be suitably defined by analytic continuation from positive
integer powers.
The latter can be done either by computing $\Box^n$ for positive integer $n$
and then analytically continuing to $n \rightarrow -1/2\nu$,
or alternatively by making use of the identity
\beq
{ 1 \over {\Box}^n } \; = \; 
{ (-1)^n \over \Gamma (n) } \, \int_0^{\infty} 
ds \, s^{n-1} \exp ( i \, s \, \Box )
\eeq
and subsequent use of the Schwinger-DeWitt representation
for the kernel $\exp ( i \, s \, \Box )$ of the massless
operator $\Box$.
Within the limited scope of this paper, we will be satisfied 
with computing the effects of positive integer powers $n$ of the
covariant d'Alembertian $\Box$,
and then analytically continue the answer to fractional $n=-1/ 2 \nu$.
In the following the above analytic continuation from positive
integer $n$ will always be understood.
\footnote{
We notice in passing that in this approach it is not obvious how to formulate
a running cosmological constant, as the d'Alembertian $\Box$ in 
$ \lambda (r) \, \int dx \sqrt{g} \, \rightarrow \,
\lambda \int dx \sqrt{g} \, 
\left ( 1 \, - \, c_{\Box} \, 
\left ( 1 / \xi^2 \Box  \right )^{ \gamma }  \right )
$ has no function of the metric left to act on ~\cite{barvi}.
This situation is not entirely surprising as, lacking derivatives, 
the effect of the $\lambda$ term is just to control the overall scale.
In pure lattice gravity the bare $\lambda$ is trivially scaled out
and does not run ~\cite{critical,ttmodes}. 
In this scenario the physical long distance cosmological constant
$\sim 1/ \xi^2$, being related to an average curvature, is
considered a physical quantity to be kept fixed as the gravitational
coupling $G(r)$ slowly evolves with scale.}

It should be stressed here that the action in Eq.~(\ref{eq:ieff_sr})
should be treated as a {\it classical} effective action, with dominant
radiative corrections at short distances ($r \ll \xi$) already automatically 
built in, and for which a restriction to generally smooth field
configurations does make some sense. 
In particular one would expect
that in most instances it should be possible, as well as meaningful, 
to neglect terms involving large numbers of derivatives of the
metric in order to compute the effects of the new contributions appearing in
the effective action.
\footnote{
Dominant contributions to the original
Feynman path integral for the underlying quantum gravity theory
are, on the other hand, presumably nowhere differentiable,
the smooth configurations having ultimately
zero measure in the gravitational functional integral \cite{feypath}.
Furthermore, issues related to causality, unitarity and positivity are
better referred to the original, local microscopic action, which presumably
shares all of these properties.}

A number of useful results can already be obtained from
the form of the effective action in Eq.~(\ref{eq:ieff_sr}).
In particular, once a specific metric is chosen,
the running of $G$ can be readily expressed in terms of the 
coordinates appropriate for that metric. 
Later in this work we will illustrate extensively this statement for the
specific, and physically relevant, case of the Robertson-Walker (RW) metric.

The next major step involves a derivation of the effective field 
equations, incorporating the running of $G$. 
As will be shown below this is not entirely straightforward, as
the variation of the non-local effective action is complicated
by the presence of a differential operator raised to a fractional power,
acting on what are rather complicated functions of the metric to begin with.
We shall therefore postpone a discussion of this 
aspect to the Appendix, which focuses on this specific
topic.

Had one {\it not} considered the action of Eq.~(\ref{eq:ieff_sr})
as a starting point for
constructing the effective theory, one would naturally be led 
(following Eq.~(\ref{eq:gbox}))
to consider the following effective field equations
\beq
R_{\mu\nu} \, - \, \half \, g_{\mu\nu} \, R \, + \, \Lambda \, g_{\mu\nu}
\; = \; 8 \pi G  \, \left( 1 + A( \Box ) \right) \, T_{\mu\nu}
\label{eq:naive_t}
\eeq
the argument again being the replacement 
$G(r) \, \rightarrow \, G \left( 1 + A( \Box ) \right) $ involving
the invariant object $\Box$.
Here, following common notation, $\Lambda$ is the scaled cosmological constant,
not to be confused with the ultraviolet cutoff.
Being manifestly covariant, these expressions at least satisfy some
of the requirements for a set of consistent field equations
incorporating the running of $G$.
The above effective field equation can then be easily re-cast in a form
similar to the classical field equations
\beq
R_{\mu\nu} \, - \, \half \, g_{\mu\nu} \, R \, + \, \Lambda \, g_{\mu\nu}
\; = \; 8 \pi G  \, {\tilde T_{\mu\nu}}
\eeq
with $ {\tilde T_{\mu\nu}} \, = \, \left( 1 + A( \Box ) \right) \, T_{\mu\nu}$
defined as an effective, or {\it gravitationally dressed}, energy-momentum tensor.
Just like the ordinary Einstein gravity case,
in general ${\tilde T_{\mu\nu}}$ might not be covariantly conserved a priori,
$\nabla^\mu \, {\tilde T_{\mu\nu}} \, \neq \, 0 $, but ultimately the
consistency of the effective field equations {\it demands} that it
be exactly conserved in consideration of the Bianchi identity satisfied
by the Riemann tensor.
The ensuing new covariant conservation law
\beq
\nabla^\mu \, {\tilde T_{\mu\nu}} \; \equiv \; 
\nabla^\mu \, \left [ \left ( 1 + A( \Box ) \right ) \, T_{\mu\nu} 
\right ] \;  = \;  0
\label{eq:continuity}
\eeq
can be then be viewed as a {\it constraint}
on ${\tilde T_{\mu\nu}}$ (or $T_{\mu\nu}$) which, for example,
in the specific case of a perfect fluid, 
will imply again a definite relationship between the density $\rho(t) $,
the pressure $p(t)$ and the RW scale factor $R(t)$, just as it does
in the standard case. 

This point is sufficiently important that we wish to elaborate on it 
further. 
In {\it ordinary} Einstein gravity the energy momentum tensor
is {\it defined} via the variation of the matter action
\beq 
\delta I_{M} \; = \; \half \int dx \sqrt{g} \, \delta g_{\mu\nu} \, T^{\mu\nu}
\label{eq:mattact}
\eeq
But when the above arbitrary variation $ \delta g_{\mu\nu} $
is taken to be a gauge variation,
\beq
\delta g_{\mu\nu} \; = \;
 \, g_{\mu\lambda} \, \partial_{\nu} \epsilon^{\lambda}
+ \, g_{\lambda\nu} \, \partial_{\mu} \epsilon^{\lambda}
+ \, \epsilon^{\lambda} \, \partial_{\lambda} \,  g_{\mu\nu}
\eeq 
integration by parts in Eq.~(\ref{eq:mattact}) immediately yields the
covariant conservation law $\nabla^\mu \, T_{\mu\nu}  \, = \, 0 $,
as a direct consequence of the gauge invariance of the matter action.

On the other hand, in the modified field equations of Eq.~(\ref{eq:naive_t}),
the object which will be required to be conserved by the
consistency of the field equations is the gravitationally
dressed energy momentum tensor, namely 
$ \left ( 1 + A( \Box ) \right ) \, T_{\mu\nu} $,
and not the original bare $T_{\mu\nu}$ itself.
Referring therefore to the original 
$T_{\mu\nu}$ as ``the energy momentum tensor'' would 
appear to be improper, since, for the consistency of the effective field
equations of Eq.~(\ref{eq:naive_t}),
the latter is no longer required to be covariantly conserved.
\footnote{
This can be illustrated further by the specific case of the perfect fluid,
for which the energy momentum tensor is usually written as
$ T_{\mu\nu} \;  = \; \left( p(t) \, + \, \rho(t) \right) u_\mu \, u_\nu
\, + \, g_{\mu \nu} \, p(t) $.
In general its covariant divergence is not zero, but consistency
of the Einstein field equations demands $\nabla^{\mu} T_{\mu\nu} =0 $,
which for the RW metric forces a definite relationship between 
$R(t)$, $\rho(t)$ and $p(t)$,
namely $ \dot{\rho}(t) + 3 \left( \rho(t) + p(t) \right)
\left( \dot{R}(t) / R(t) \right) \; = \; 0 $, irrespective of the equation
of state relating $\rho$ to $p$.
In the effective field equations of Eq.~(\ref{eq:naive_t}) 
the perfect fluid form for $T_{\mu\nu}$ can still be used (at it still
satisfies all the original symmetry requirements), but the covariant
conservation law has the new form displayed in Eq.~(\ref{eq:continuity}), which
imposes a {\it new} constraint on the scale factor
$R(t)$, as well as on the underlying $\rho(t)$ and $p(t)$.}
In a sense, the effective field equations of Eq.~(\ref{eq:naive_t}) can be seen
simply as a consequence of having changed the expression in
Eq.~(\ref{eq:mattact}) to
\beq 
\delta I_{M}^{\; '} \; = \; \half \int dx \sqrt{g} \; \delta g_{\mu\nu} \, 
\left( 1 + A( \Box ) \right) \, T^{\mu\nu}
\label{eq:mattactmod}
\eeq
Let us make a few more comments regarding the above effective field equations,
in which we will set the cosmological constant $\Lambda=0$ from now on.
One simple observation is that the trace equation only involves the (simpler)
scalar d'Alembertian, acting on the trace of the energy-momentum tensor
\beq
R \; = \; 8 \pi G  \, \left( 1 + A( \Box ) \right) \, T_{\mu}^{\;\;\mu}
\label{eq:naive_tt}
\eeq
Furthermore, to the order one is working here, the
above effective field equations should be equivalent to
\beq
\left( 1 \, - \, A( \Box )  \, + \, O(A( \Box )^2) \right) 
\left( R_{\mu\nu} - \half \, g_{\mu\nu} \, R \right)
\; = \; 8 \pi G \, T_{\mu\nu}
\label{eq:naive_r}
\eeq
where the running of $G$ has been moved over to the ``gravitational'' side.
Indeed it has recently been claimed ~\cite{barvi} that equations
similar to the above effective field equations
(at least for positive integer power $n$, including the classical case $n=0$)
can be derived from a nonlocal extension of the
Einstein-Hilbert action. 
In the classical case ($A(\Box)=0$) one writes a new non-local action
\beq
I \; = \; { 1 \over 16 \pi G } \int dx \sqrt{g} 
\left( R^{\mu\nu} \, - \, \half \, g^{\mu\nu} \, R \right) { 1 \over \Box }
R_{\mu\nu}
\eeq
whose variation, it is argued, gives the correct field equations
up to curvature squared terms
\beq
\sqrt{g} \left( R^{\mu\nu} - \half \, g^{\mu\nu} \, R 
\; + \; O \left( R_{\mu\nu}^2 \right) \right)  \; = \; 0
\eeq
For non-vanishing $A(\Box)$ the above construction can 
then be generalized to
\beq
I \; = \; { 1 \over 16 \pi G } \int dx \sqrt{g} 
\left( R^{\mu\nu} - \half \, g^{\mu\nu} \, R \right) 
\left( 1 - A( \Box ) \, + \, O(A( \Box )^2) \right) { 1 \over \Box }
R_{\mu\nu}
\eeq
whose variation can now be shown to give
\beq
\sqrt{g} 
\left( 1 - A( \Box ) \, + \, O(A( \Box )^2) \right)
\left( R_{\mu\nu} - \half \, g_{\mu\nu} \, R \right)
\; + \; O \left( R_{\mu\nu}^2 \right) \; = \; 0
\eeq
and which would coincide with the previously proposed effective
field equations, again up to higher order curvature terms.

\vskip 30pt
\newsection{Covariant d'Alembertian on Scalar Functions}
\hspace*{\parindent}

As a first step in solving the new set of effective field equations, consider
first the {\it trace}
of the field equation in Eq.~(\ref{eq:naive_tt}),
written as 
\beq
\left( 1 \, - \, A( \Box ) \, + \, O ( A( \Box )^2 ) \right) \, R 
\; = \; 8 \pi G  \, T_{\mu}^{\;\;\mu}
\label{eq:naive_rt}
\eeq
where $R$ is the scalar curvature.
Here we have made the choice to move the operator $A(\Box)$ over
on the gravitational side, so that it now acts on functions of
the metric only, using the binomial expansion of
$1/( 1 \, + \, A( \Box ) )$.
A discussion of the full tensor equations and their added
complexity will be postponed to the next section.
To proceed further, one needs to compute the effect of $A(\Box)$
on the scalar curvature.
The d'Alembertian operator acting on scalar functions $S(x)$ is given by
\beq
{1 \over \sqrt{g} } \, \partial_\mu \, g^{\mu\nu} \sqrt{g} \, \partial_\nu \, S(x)
\eeq
and for the RW metric, acting on functions of $t$ only, one obtains
a fairly simple result in terms of the scale factor $R(t)$
\beq
- { 1 \over R^3(t) } \, { \partial \over \partial t } \left [ 
R^3(t) { \partial \over \partial t } \right ] \; F(t)
\eeq
As a next step one computes the action of $\Box$ on the scalar curvature $R$, which
gives
\beq
-6\,\left( -2\,k\,\ddot{R}(t) - 5\,{\dot{R}}^2(t)\,\ddot{R}(t) + R(t)\,{\ddot{R}}^2(t) + 
      3\,R(t)\,\dot{R}(t)\,R^{(3)}(t) + R^2(t)\,R^{(4)}(t) \right) / R^3(t)
\eeq
and then $\Box^2$ on $R$ which gives
\bea
& 6 \, ( -6\,k\, {\dot{R}}^2(t) \, {\ddot{R}(t)} - 15\, {\dot{R}}^4(t) \,\ddot{R}(t) + 
6\,k\,R(t)\, {\ddot{R}}^2(t) + 45\,R(t)\, {\dot{R}}^2(t) \, {\ddot{R}}^2(t) - \nonumber \\
&    12\, R^2(t) \, {\ddot{R}}^3(t) + 6\,k\,R(t)\,\dot{R}(t)\, R^{(3)} (t) + 
15\,R(t)\, {\dot{R}}^3(t) \, R^{(3)} (t) 
- 41\, R^2(t) \, {\dot{R}(t)} \, {\ddot{R}(t)} \, R^{(3)} (t)+
\nonumber \\
&    5\, R^3(t) \, {R^{(3)}}^2(t) - 2\,k\, R^2(t) \, R^{(4)}(t) - 
9\, R^2(t) \, {\dot{R}}^2(t) \, R^{(4)}(t) + 7\, R^3(t) \, {\ddot{R}(t)} \, R^{(4)} (t) +
\nonumber \\
&    4\, R^3(t) \, {\dot{R}(t)} \, R^{(5)} (t) + {R(t)}^4 \, R^{(6)}(t) 
) / {R(t)}^5
\eea
etc. It should already become clear at this point that
the computed expressions are rapidly becoming quite complicated.
Nevertheless some of the higher order terms can, for example, be interpreted as
higher derivative
curvature contributions, since for Riemann squared, Ricci squared and scalar
curvature squared, one has respectively
\beq
R_{\mu\nu\lambda\sigma} \, R^{\mu\nu\lambda\sigma}
 \; = \; 12 \left( 
k^2 \, + \, 2 k {\dot{R}}^2(t) \, + \, {\dot{R}}^4(t) \, + \, R^2(t) {\ddot{R}}^2(t)
\right) / {R(t)}^4
\eeq
\beq
R_{\mu\nu} \, R^{\mu\nu} \; = \; 12 \left(
k^2 \, + \, 2 k {\dot{R}}^2(t) \, + \, k {R(t)} {\ddot{R}(t)}
\, + \, {\dot{R}}^4(t) \, + \, R^2(t) {\ddot{R}}^2(t)  
\, + \, {R(t)} {\dot{R}}^2(t)  {\ddot{R}(t)} \right)
/ {R(t)}^4
\eeq
\beq
R^2 \; = \; 36 \left( k \, + \, {\dot{R}}^2(t) \, + \, R(t) \ddot{R}(t) \right)^2
/ {R(t)}^4
\label{eq:r2}
\eeq
with
\beq
R_{\mu\nu\lambda\sigma} \, R^{\mu\nu\lambda\sigma} 
\, - \, \sixth \, R_{\mu\nu} \, R^{\mu\nu} \, - \, \half \, R^2 \; = \; 0 
\eeq
for arbitrary scale factor $R(t)$.
But in the following we will just simply set $R(t) = R_0 \, t^{\alpha}$,
in which case 
\beq
R_{\mu\nu\lambda\sigma} \, R^{\mu\nu\lambda\sigma}
 \; = \; { 12 \, \alpha^2 ( 2 \, \alpha^2 - 2 \, \alpha +1 ) \over t^4 } 
\eeq
\beq
R_{\mu\nu} \, R^{\mu\nu}
 \; = \; { 12 \, \alpha^2 ( 3 \, \alpha^2 - 3 \, \alpha +1 ) \over t^4 } 
\eeq
\beq
R^2  \; = \; { 36 \, \alpha^2 ( 2 \, \alpha - 1 )^2  \over t^4 } 
\label{eq:r22}
\eeq
and for the scalar curvature (here allowing for $k \neq 0$, see
Eq.~(\ref{eq:scalar}) in Appendix A)
\beq
6 \left( 
{ k \over R_0^2 \, t^{2\alpha} } +
{ \alpha \left( -1 + 2 \, \alpha \right) \over t^2 } \right)
\eeq
Acting with $\Box^n$ on the above scalar curvature now gives for $k=0$
\beq
6\,\alpha \,\left( -1 + 2\,\alpha  \right) t^{-2}
\eeq

\beq
36\,\left( -1 + \alpha  \right) \,\alpha \,\left( -1 + 2\,\alpha  \right) t^{-4}
\eeq

\beq
144\,\left( -1 + \alpha  \right) \,\alpha \,\left( -1 + 2\,\alpha  \right) \,
    \left( -5 + 3\,\alpha  \right) t^{-6}
\eeq

\beq
864\,\left( -1 + \alpha  \right) \,\alpha \,\left( -1 + 2\,\alpha  \right) \,
    \left( -7 + 3\,\alpha  \right) \,\left( -5 + 3\,\alpha  \right) t^{-8}
\eeq
for $n=0$, 1, 2, and 3 respectively, and therefore for arbitrary power $n$ 
\beq
c_n \, 6 \, \alpha \, \left( -1 + 2\,\alpha  \right) t^{-2-2n}
\eeq
with the coefficient $c_n$ given by
\beq
c_n \; = \; 4^n 
{ \Gamma ( n + 1 ) \Gamma ( { 3 \alpha - 1 \over 2  } ) \over
\Gamma ( { 3 \alpha - 1 \over 2 } - n ) }
\eeq
Here use has been made of the relationship
\beq
\left( { d \over d \, z } \right)^{\alpha} \, \left( z \, - \, c \right)^{\beta}
\; = \; { \Gamma ( \beta + 1 ) \over \Gamma ( \beta - \alpha + 1 ) }
\, \left( z \, - \, c \right)^{\beta-\alpha}
\eeq
to analytically continue the above expressions to negative
fractional $n$ \cite{frac} .
For $n=-1/2\nu$ the correction on the scalar curvature
term $R$ is therefore of the form
\beq
\left( 1 - c_\nu \, (t/\xi)^{1/\nu} \right) \, \cdot \, 6 
\, \alpha \, \left( -1 + 2\,\alpha  \right) t^{-2}
\eeq
with 
\beq
c_{\nu} \; = \; 2^{-{1 \over \nu}} 
{ \Gamma ( 1 - {1 \over 2 \nu} ) \Gamma ( { 3 \alpha - 1 \over 2 } ) \over
\Gamma ( { 3 \alpha - 1 \over 2 } + {1 \over 2 \nu } ) }
\label{eq:cr}
\eeq
In particular for $\alpha=2/3$ (the classical value for a pressureless
perfect fluid) and $\nu=1/3$ one has
\beq
c_{\nu} \; = \; 2^{-3} 
{ \Gamma ( - {1 \over 2} ) \Gamma ( { 1 \over 2 } ) \over
\Gamma ( 2 ) }  \; = \; - { \pi \over 4 }
\eeq
whereas, for example,  for $\alpha=1/2$ and $\nu=1/3$ one obtains 
$c_{\nu} = - \sqrt{\pi} \, \Gamma ( \frac{5}{4} ) / \Gamma ( \frac{7}{4} )$.
Putting everything together, one then obtains for the trace
part of the effective field equations
\beq 
\left ( 
1 - c_{\Box} \, c_{\nu} \, \left ( { t \over \xi } \right )^{1/\nu} 
\, + \, O \left (  ( t / \xi )^{2 / \nu}  \right ) 
\right ) \, 
{ 6 \, \alpha \, \left( 2\,\alpha -1 \right)  \over t^2 } 
\; = \; 8 \pi G \, \rho (t)
\eeq
The new term can now be moved back over to the matter side
(since the correction is assumed to be small), in accordance
with the structure of the original effective field equations
Eqs~.(\ref{eq:naive_t}) and (\ref{eq:naive_tt}), and
thus avoids the problem of having to deal with the binomial
expansion of $1/(1 \, + \, A(\Box))$. 
One then has
\beq 
{ 6 \, \alpha \, \left( 2\,\alpha -1 \right) \over t^2 }
\; = \; 8 \pi G \, \left ( 1 + c_{\Box} \, c_{\nu}
\, \left ( { t \over \xi } \right )^{1/\nu}
\, + \, O \left (  ( t / \xi )^{2 / \nu}  \right ) 
\right ) \, \rho (t)
\eeq
which is the RW metric form of Eq.~(\ref{eq:naive_tt}).
If one assumes for the matter density
$\rho(t) \sim \rho_0 \, t^{\beta}$, then matching powers when
the new term starts to take over at larger distances gives the first result
\beq
\beta = -2 - 1/\nu 
\label{eq:beta}
\eeq 
Thus the density decreases {\it faster} in time than the classical value
$(\beta =-2)$ would indicate.
The expansion appears therefore to be accelerating, but before reaching such
a conclusion one needs to determine the time dependence of the scale factor
$R(t)$ (or $\alpha$) as well.

One might be troubled by the fact that some of the Gamma functions
appearing in the expression for $c_{\nu}$ can
diverge for specific choices of $\nu$, e.g. when
$\nu=1/2(n+1)$ as in Eq.~(\ref{eq:cr}) for $n$ integer.
But further thought reveals that this is not necessarily a concern here,
as the coefficient $c_{\nu}$ actually has to be divided
out and then multiplied by $c_\xi$ (which, as discussed in
the introduction and in ~\cite{ttmodes}, is expected to be a number of order one)
to get the correct magnitude for the correction.
One has therefore
\beq
c_{\Box} \, c_{\nu} \; = \; c_\xi 
\eeq
so that the correction eventually ends up as $(1 + c_\xi (t/\xi)^{1/\nu})$,
as it should, in accordance with Eq.~(\ref{eq:grun})
for $G(r)$ (the ``t'' here is like ``r'' there).

Having completed the calculation of the quantum correction term
acting on the scalar curvature, as in Eq.~(\ref{eq:naive_rt}), one
can alternatively pursue the following exercise in order to check
the overall consistency of the approach.
Consider $\Box^n $ acting on $T_{\mu}^{\;\; \mu} = - \rho(t) $ instead,
as in the trace of the effective field equation Eq.~(\ref{eq:naive_tt})
\beq
R \; = \; 8 \pi G  \, \left( 1 + A( \Box ) \right) \, T_{\mu}^{\;\;\mu}
\eeq
for $\Lambda=0$ and $p(t)=0$.
For $\rho(t) = \rho_0 \, t^{\beta}$ and $R(t) = R_0 \, t^{\alpha}$
one finds in this case
\beq
\Box^n \left( - \rho(t) \right) \; \rightarrow \; 4^n (-1)^{n+1}  
{ \Gamma ( {\beta \over 2} + 1 ) 
  \Gamma ( { \beta + 3 \alpha + 1 \over 2 } ) 
\over
  \Gamma ( {\beta \over 2} + 1 - n ) 
  \Gamma ( { \beta + 3 \alpha + 1 \over 2 } - n  )  } \; \rho_0 \, t^{\beta - 2n}
\eeq
which again implies $\beta = -2 - 1/\nu$ as in Eq.~(\ref{eq:beta})
for large(r) times, when
the quantum correction starts to become important (since the left hand side 
of Einstein's equation always goes like $1 / t^2 $, no matter what the value
for $\alpha$ is, at least for k=0).

\newpage

\vskip 30pt
\newsection{Covariant d'Alembertian on Tensor Functions}
\hspace*{\parindent}

Next we will examine the full effective field equations (as opposed
to just their trace part) of Eq.~(\ref{eq:naive_t}) with $\Lambda=0$,
\beq
R_{\mu\nu} \, - \, \half \, g_{\mu\nu} \, R \, 
\; = \; 8 \pi G  \, \left( 1 + A( \Box ) \right) \, T_{\mu\nu}
\eeq
Here the d'Alembertian operator
\beq
\Box \; = \; g^{\mu\nu} \nabla_\mu \nabla_\nu 
\eeq
acts on a second rank tensor,
\bea
\nabla_{\nu} T_{\alpha\beta} \, = \, \partial_\nu T_{\alpha\beta} 
- \Gamma_{\alpha\nu}^{\lambda} T_{\lambda\beta} 
- \Gamma_{\beta\nu}^{\lambda} T_{\alpha\lambda} \, \equiv \, I_{\nu\alpha\beta}
\nonumber
\eea
\beq 
\nabla_{\mu} \left( \nabla_{\nu} T_{\alpha\beta} \right)
= \, \partial_\mu I_{\nu\alpha\beta} 
- \Gamma_{\nu\mu}^{\lambda} I_{\lambda\alpha\beta} 
- \Gamma_{\alpha\mu}^{\lambda} I_{\nu\lambda\beta} 
- \Gamma_{\beta\mu}^{\lambda} I_{\nu\alpha\lambda} 
\eeq
and would thus seem to require the calculation of 1920 terms,
of which fortunately many vanish by symmetry.
Next assume that $T_{\mu\nu}$ has the perfect fluid form, for which one 
obtains 
\bea
\left( \Box \, T_{\mu\nu} \right )_{tt} \; & = & \; 
6 \, \left [ \rho (t) \, + \, p(t) \right ]
\, \left ( { \dot{R}(t) \over R(t) } \right )^2
\, - \, 3 \, \dot{\rho}(t) \,  { \dot{R}(t) \over R(t) }
\, - \, \ddot{\rho}(t) 
\nonumber \\
\left( \Box \, T_{\mu\nu} \right )_{rr} \; & = & \; 
{ 1 \over 1 \, - \, k \, r^2 } \left \{
2 \, \left [ \rho (t) \, + \, p(t) \right ] \, \dot{R}(t)^2 
\, - \, 3 \, \dot{p}(t) \, R(t) \, \dot{R}(t) 
\, - \, \ddot{p}(t) \, R (t)^2  \right \}
\nonumber \\
\left( \Box \, T_{\mu\nu} \right )_{\theta\theta} \; & = & \; 
r^2 \, ( 1 \, - \, k \, r^2 ) \, 
\left( \Box \, T_{\mu\nu} \right )_{rr}
\nonumber \\
\left( \Box \, T_{\mu\nu} \right )_{\varphi\varphi} \; & = & \; 
r^2 \, ( 1 \, - \, k \, r^2 ) \, 
\sin^2 \theta \, \left( \Box \, T_{\mu\nu} \right )_{rr}
\label{eq:boxont}
\eea
with the remaining components equal to zero.
Note that a non-vanishing pressure contribution is generated in the effective
field equations, even if one assumes initially a pressureless fluid, $p(t)=0$.
As before, repeated applications of the d'Alembertian $\Box$ to the above expressions leads
to rapidly escalating complexity (for example, eighteen distinct 
terms are generated by $\Box^2$ for each of the above contributions),
which can only be tamed by introducing some further simplifying assumptions.
In the following we will therefore assume that 
$T_{\mu\nu}$ has the perfect fluid form appropriate
for non-relativistic matter, with a power law behavior for
the density, $\rho(t) = \rho_0 \, t^\beta$, and $p(t)=0$. 
Thus all components of $T_{\mu\nu}$ vanish in the fluid's
rest frame, except the $tt$ one, which is simply $\rho(t)$.
Setting $k=0$ and $R(t) = R_0 \, t^\alpha $ one then finds
\bea
\left( \Box \, T_{\mu\nu} \right )_{tt} \; & = & \; 
\left( 6 \, \alpha^2 - \beta^2 - 3 \, \alpha \, \beta  + \beta \right) \, \rho_0 \,
t^{\beta - 2 }
\nonumber \\
\left( \Box \, T_{\mu\nu} \right )_{rr} \; & = & \; 
2 R_0^2 \, t^{2 \alpha} \alpha^2  \, \rho_0 \, t^{\beta -2 }
\eea
which again shows that the $tt$ and $rr$ components get mixed by the
action of the $\Box$ operator, and that a 
non-vanishing $rr$ component gets generated,
even though it was not originally present.

Higher powers of the d'Alembertian $\Box$ acting on 
$T_{\mu\nu}$ can then be computed as well,
but it is easier to introduce the slightly more general
auxiliary diagonal tensor $V_{\mu\nu}$ with
components $V_{tt} = \rho_0 \, t^\beta$, $V_{rr} = \rho_1 \, t^\gamma$,
$V_{\theta\theta} = r^2 \, V_{rr} $ and
$V_{\varphi\varphi} = r^2 \, \sin^2 \theta \, V_{rr}$, with $\gamma$ an
arbitrary power.
One then finds
\bea
\left( \Box \, V_{\mu\nu} \right )_{tt} \; & = & \; 
\left( 6 \, \alpha^2 - \beta^2 - 3 \, \alpha \, \beta  + \beta \right) \, \rho_0 \,
t^{\beta - 2 } 
\; + \; { 6 \, \alpha^2 \over R_0^2 \, t^{2 \alpha} } \, \rho_1 \, t^{\gamma-2}
\nonumber \\
\left( \Box \, V_{\mu\nu} \right )_{rr} \; & = & \; 
2 R_0^2 \, t^{2 \alpha} \alpha^2  \, \rho_0 \, t^{\beta -2 }
\; + \; 
\left( 4 \, \alpha^2 + \alpha \, ( \gamma - 2 ) - \gamma \, ( \gamma - 1 ) 
\right) \, \rho_1 \, t^{\gamma -2 }
\eea
as well as
\bea
\left( \Box \, V_{\mu\nu} \right )_{\theta\theta} \; & = & \; 
r^2 \, \left( \Box \, V_{\mu\nu} \right )_{rr} 
\nonumber \\
\left( \Box \, V_{\mu\nu} \right )_{\varphi\varphi} \; & = & \; 
r^2 \, \sin^2 \theta \, \left( \Box \, V_{\mu\nu} \right )_{rr}
\eea
and zero for the remaining components.
The above expressions can then be used conveniently to generate $\Box^n$
acting on $T_{\mu\nu}$ to any desired power $n$.
But since the problem at each step involves a two by two matrix acting
on the energy momentum tensor, it would seem rather complicated to get
a closed form solution for arbitrary $n$.
But a comparison with the left hand (gravitational) side of the effective
field equation, which always behaves like $\sim 1 / t^2 $ for $k=0$, shows
that in fact a solution can only be achieved at order $\Box^n$ 
provided the exponent $\beta$ satisfies $\beta = -2 + 2 n $, or
since $n=-1/( 2 \nu )$,
\beq
\beta \; = \; - 2 \, - \, 1 / \nu 
\label{eq:beta1}
\eeq
as was found previously from the trace equation,
Eqs.~(\ref{eq:naive_tt}) and (\ref{eq:beta}).
As a result one obtains a much simpler set of expressions, which now read
\beq
\left( \Box \, T_{\mu\nu} \right )_{tt} \; \rightarrow \; 
6 \, \alpha^2 \, \rho_0 \, t^{- 2 } 
\eeq

\beq
\left( \Box^2 \, T_{\mu\nu} \right )_{tt} \; \rightarrow \; 
12 \, \alpha^2 \, ( \alpha -1 ) \, ( 4 \alpha + 1) \, \rho_0 \, t^{- 2 } 
\eeq

\beq
\left( \Box^3 \, T_{\mu\nu} \right )_{tt} \; \rightarrow \; 
48 \, \alpha^2 \, ( \alpha -1 ) \, ( 4 \alpha + 1) \,
( 2 \, \alpha^2 - 3 \, \alpha - 3 ) \, \rho_0 \, t^{- 2 } 
\eeq

\beq
\left( \Box^4 \, T_{\mu\nu} \right )_{tt} \; \rightarrow \; 
96 \, \alpha^2 \, ( \alpha -1 ) \, ( 4 \alpha + 1) \,
( 2 \, \alpha^2 - 3 \, \alpha - 3 ) \,
( 4 \, \alpha^2 - 9 \, \alpha - 15 ) \, \rho_0 \, t^{- 2 } 
\eeq

\beq
\left( \Box^5 \, T_{\mu\nu} \right )_{tt} \; \rightarrow \; 
768 \, \alpha^2 \, ( \alpha -1 ) \, ( 4 \alpha + 1) \,
( 2 \, \alpha^2 - 3 \, \alpha - 3 ) \,
( 4 \, \alpha^2 - 9 \, \alpha - 15 ) \, 
(  \alpha^2 - 3 \, \alpha - 7 ) \, \rho_0 \, t^{- 2 } 
\eeq
etc., here for powers $n=1$ to $n=5$ respectively, and with
$\beta$ changing with $n$ in accordance with Eq.~(\ref{eq:beta1}).
For general $n$ one can then write 
\beq
\left( \Box^n \, T_{\mu\nu} \right )_{tt} \; \rightarrow \; 
c_{tt} ( \alpha, \nu ) \, \rho_0 \, t^{- 2 } 
\eeq
and similarly for the $rr$ component
\beq
\left( \Box^n \, T_{\mu\nu} \right )_{rr} \; \rightarrow \; 
c_{rr} ( \alpha, \nu ) \, R_0^2 \, t^{2 \alpha} \, \rho_0 \, t^{- 2 } 
\eeq
But remarkably (see also Eq.~(\ref{eq:boxont}) )
one finds for the two coefficients the simple identity
\beq
c_{rr} (\alpha, \nu ) \; = \; \third \, c_{tt} ( \alpha, \nu)
\label{eq:crr}
\eeq
as well as $ c_{\theta\theta} = r^2 \, c_{rr} $ and
$ c_{\varphi\varphi} = r^2 \, \sin^2 \theta \, c_{rr} $.
Then for large times, when the quantum correction starts to become important,
the $tt$ and $rr$ field equations reduce to
\beq
3 \, \alpha^2 \, t^{-2} \; = \; 8 \pi G \,
c_{tt} ( \alpha, \nu ) \, \rho_0 \, t^{- 2 } 
\eeq
and 
\beq
- \; \alpha \, ( 3 \, \alpha \, - \, 2 ) \, R_0^2 \, t^{2 \alpha - 2 } 
\; = \; 8 \pi G \,
c_{rr} ( \alpha, \nu ) \, R_0^2 \, t^{2 \alpha} \, \rho_0 \, t^{- 2 } 
\eeq
respectively.
But the identity $ c_{rr} \; = \; \third \, c_{tt} $
implies, simply from the consistency of the $tt$ and $rr$ effective
field equations at large times,
\beq
{ c_{rr} ( \alpha, \nu)  \over c_{tt} (\alpha, \nu ) } \; \equiv \; \third 
\; = \; - \, { 3 \alpha - 2  \over 3 \alpha }
\eeq
whose only possible solution finally gives the second sought-for result, namely
\beq
\alpha \; = \; { 1 \over 2 }
\label{eq:alpha}
\eeq
For the specific value of $\alpha=\half$ one can then show
that the coefficients $c_{tt}$ obey the recursion relation
\beq
(c_{tt})_n \; = \; - ( 4 \, n^2 \, - \, 7 n + 1) \, (c_{tt})_{n-1}
\eeq 
with initial condition $ (c_{tt})_{n=1} = 3/2 $.
Consequently a closed form expression for $c_{tt}$
and $c_{rr} = c_{tt}/3$
can be written down, either in terms of the Pochhammer symbol
$(x)_n = x (x+1) \dots (x+n-1) = \Gamma(x+n)/ \Gamma(x)$, or more
directly in terms of ratios of Gamma functions as
\beq
c_{tt} ( \alpha = \half, n = -1/2 \nu ) \; = \;
 3 \, {\left( -1 \right) }^{n+1} \, 2^{-3 + 2 \, n} \,
    \frac{
    \Gamma (\frac{1 - {\sqrt{33}}}{8} + n)
    \Gamma (\frac{1 + {\sqrt{33}}}{8} + n) } {
    \Gamma (\frac{9 - {\sqrt{33}}}{8}) \, 
    \Gamma (\frac{9 + {\sqrt{33}}}{8}) }
\eeq 
Still, the above expression does not seem to be particularly
illuminating at this point, except for providing an explicit proof
that the coefficients
$c_{tt}$ and $c_{rr}$ exist and are finite for specific values
of $n$, such as $n=-1/2 \nu = -3/2$.

One might worry at this point whether the above solution is consistent
with covariant energy conservation.
With the assumed form for $T_{\mu\nu}$
it is easy to check that energy conservation yields for the $t$ component
\beq
\left ( \nabla^\mu \, \left ( \Box^n \, T_{\mu\nu} \right ) \right )_t 
\; \rightarrow \; - \, \left ( 
( 3 \, \alpha \, + \, \beta \, + \, 1 / \nu ) \, c_{tt} \, + \,
3 \, \alpha \, c_{rr} 
\right ) \, \rho_0 \, t^{\beta + 1 / \nu - 1} \; = \; 0
\eeq
when evaluated for $n=-1 / 2 \nu$, and zero for the remaining three spatial
components.
But from the solution for the matter density $\rho(t)$ at large times one
has $\beta = - 2 - 1 / \nu $, so the above zero condition gives again
$ c_{rr} / c_{tt} = - ( 3 \alpha - 2 )/ 3 \alpha $, exactly 
the same relationship previously implied by the consistency of the
$tt$ and $rr$ field equations.

In conclusions the values for $\alpha= 1/2 $ of Eq.~(\ref{eq:alpha}) and 
$\beta =-2 -1/ \nu $ of Eq.~(\ref{eq:beta1}), determined from
the effective field equations at large times, are found to be consistent
with {\it both} the field equations {\it and} covariant energy conservation.
More importantly, the above solution is also consistent with what was found
previously by looking at the trace of the effective field equations,
Eq.~(\ref{eq:naive_tt}), which also implied the result
$\beta=-2 -1/ \nu$, Eq.~(\ref{eq:beta}).

Together these results imply that for sufficiently large times
the scale factor $R(t)$ behaves as
\beq
R(t) \; \sim \; t^{\alpha} \; \sim \; t^{1/2}
\eeq
and the density $\rho(t)$ as
\beq
\rho (t) \; \sim \; t^{\beta} \; \sim \;
t^{- 2 - 1 / \nu } \; \sim \; 
\left ( R(t) \right )^{- 2 \, ( 2 + 1 / \nu ) }
\eeq 
Thus the density decreases significantly faster in time
than the classical value ($ \rho (t) \sim t^{-2}$),
again a signature of an accelerating expansion at later times.

It is amusing to note that the vacuum polarization term we have
been discussing so far behaves very much like a positive pressure term, as 
should already have been clear from the fact that the covariant
d'Alembertian $g^{\mu\nu} \nabla_\mu \nabla_\nu $
causes, in the RW metric case, a mixing
of the $tt$ and $rr$ components in the field equations.
Furthermore, within the classical FRW model, 
the value $\alpha=1/2$ corresponds to an equation-of-state parameter
$\omega=1/3$ in Eq.~(\ref{eq:alphaomega}), with
\beq
\alpha \; = \; { 2 \over 3 ( 1 \, + \, \omega ) }
\eeq
where $p(t) = \omega \, \rho(t)$, and which is therefore characteristic
of radiation.
Thus one can visualize the covariant gravitational vacuum polarization
contribution as behaving to some extent like classical radiation,
here in the form of a dilute gas of virtual gravitons.

\vskip 30pt
\newsection{Conclusions}
\hspace*{\parindent}

The main results of this paper are the effective field equations
of Eq.~(\ref{eq:naive_t}), 
\beq
R_{\mu\nu} \, - \, \half \, g_{\mu\nu} \, R \, + \, \Lambda \, g_{\mu\nu}
\; = \; 8 \pi G  \, \left( 1 + A( \Box ) \right) \, T_{\mu\nu}
\eeq
their trace in (Eq.~(\ref{eq:naive_tt})), and
the solution for the trace and full equations for the specific
case of the RW metric and $\Lambda=0$ outlined in Sections 3 and 4,
respectively.

The combined results for the density $\rho(t) \sim \rho_0 \, t^{\beta}$,
namely $\beta = - 2 - 1 / \nu $ for large times 
(Eqs.~(\ref{eq:beta}) and (\ref{eq:beta1})), and for the
scale factor $R(t) \sim R_0 \, t^{\alpha}$, namely $\alpha=1/2$ 
(Eq.~(\ref{eq:alpha})) again for large times,
imply that for $\Lambda=0$ and for sufficiently
large times the density falls off as
\beq
\rho (t) \; \sim \; t^{- 2 - 1 / \nu } \; \sim \; 
\left ( R(t) \right )^{- 2 (2 + 1 / \nu )  }
\eeq 
Thus the matter density decreases significantly faster in time
than predicted by the classical value ($ \rho (t) \sim t^{-2}$),
a signature of an accelerating expansion at later times.

Within the Friedmann-Robertson-Walker (FRW) framework
the gravitational vacuum polarization term behaves in many ways
like a positive pressure term.
The value $\alpha=1/2$ corresponds to $\omega=1/3$ in
Eq.~(\ref{eq:alphaomega}), 
\beq
\alpha \; = \; { 2 \over 3 ( 1 \, + \, \omega ) }
\eeq
where we have taken the pressure and density to be related by 
$p(t) = \omega \, \rho(t)$, and which is therefore characteristic of radiation.
One can therefore visualize
the gravitational vacuum polarization contribution
as behaving like ordinary radiation, in the form of a dilute virtual
graviton gas.
It should be emphasized though that the relationship between density $\rho (t)$
and scale factor $R(t)$ is very different from the classical case.

The results of Section 4 show that the effective Friedmann equations
for a universe filled with non-relativistic matter ($p$=0)
have the following appearance
\bea
{ k \over R^2 (t) } \, + \,
\left ( { \dot{R}(t) \over R(t) } \right )^2  
& \; = \; & { 8 \pi G(t) \over 3 } \, \rho (t) \, + \, \third \, \Lambda
\nonumber \\
& \; = \; & { 8 \pi G \over 3 } \, \left [ 
1 \, + \, c_\xi \, ( t / \xi )^{1 / \nu} 
\, + \, O \left (  ( t / \xi )^{2 / \nu}  \right ) \right ]  \, \rho (t)
\, + \, \third \, \Lambda 
\label{eq:fried_tt}
\eea
\bea
{ k \over R^2 (t) } \, + \,
\left ( { \dot{R}(t) \over R(t) } \right )^2  
\, + \, { 2 \ddot{R}(t) \over R(t) } 
& \; = \; & - \, { 8 \pi G \over 3 } \, \left [ c_\xi \, ( t / \xi )^{1 / \nu} 
\, + \, O \left (  ( t / \xi )^{2 / \nu}  \right ) \right ] \, \rho (t) 
\, + \, \Lambda
\label{eq:fried_rr}
\eea
with the running $G$ appropriate for the RW metric appearing explicitly in
the first equation,
\footnote{
Corrections to the above formulae are expected to be fixed by higher order
terms in the renormalization group $\beta$-function.
In the vicinity of the fixed point at $G_c$ one writes
\bea
\beta (G) \; = \; \beta_0 \, \left ( G \, - \, G_c \right ) 
\, + \, \beta_1 \, \left ( G \, - \, G_c \right )^2 \, + \, \cdots
\nonumber
\eea
and obtains then by integration
\bea
\xi^{-1} \; = \; C_m \, \Lambda \, 
\left \vert \, \exp \, \right \{ 
- \, \int^{G(\Lambda)} { d G' \over \beta ( G' ) }
\left \} \, \right \vert 
\; \mathrel{\mathop\sim_{G ( \Lambda )\rightarrow G_c }} \;
C_m \, \Lambda \, | \, G ( \Lambda ) \, - \, G_c |^{ - 1 / \beta ' (G_c) }
\nonumber
\label{eq:xi_cont}
\eea
with an exponent $\nu$ given by 
$\beta ' (G_c) \, = \, \beta_0 \, = \, - 1/ \nu$, $C_m$ a numeric constant
and $\Lambda$ the ultraviolet cutoff.
After replacing $\Lambda \rightarrow 1/r $ and $G(\Lambda) \rightarrow G(r) $
one finds for the scale dependence of $G$
\bea
G(r) \; = \; G_c \left [
1 \, + \, { 1 \over (1 + \beta_1 G_c \nu ) \, C_m^{1 / \nu} } 
\left ( { r \over \xi } \right )^{1 / \nu}
- \, { \beta_1 G_c \nu \over (1 + \beta_1 G_c \nu )^2  \, C_m^{2 / \nu} } 
\left ( { r \over \xi } \right )^{2 / \nu} \, + \, \cdots \right ]
\nonumber
\eea
Note that $\beta_0 = - 1/ \nu < 0$, and that for $\beta_1 < 0 $ the
second correction term is positive as well.
If one restricts oneself to the lowest order term, valid in the vicinity
of the ultraviolet fixed point, then for a given static source of
mass $M$ one has for the gravitational potential the additional contribution
$ \delta V(r) \, \sim \, ( 2 M G / \xi^3 ) \, r^2 $ for $\nu=1/3$, 
as discussed in \cite{ttmodes}.}
\beq
G(t) \; = \; G \, \left [
1 \, + \, c_\xi \, ( t / \xi )^{1 / \nu} 
\, + \, O \left (  ( t / \xi )^{2 / \nu}  \right ) \right ]
\eeq
and used, in the second equation, the result 
$ c_{tt} \; = \; 3 \, c_{rr} $ of Eq.~(\ref{eq:crr}).
We have also restored the cosmological constant term, with
a scaled cosmological constant $\Lambda \sim 1 / \xi^2 $.
One can therefore sensibly talk about an effective density
\beq
\rho_{eff} (t) \; = \; { G(t) \over G } \, \rho (t) \; = \;
\left [ 1 \, + \, c_\xi \, ( t / \xi )^{1 / \nu}  \, + \, \cdots \right ]
\, \rho (t) 
\label{eq:rho_eff}
\eeq
and an effective pressure
\beq
p_{eff} (t) \; = \; 
{ 1 \over 3 } \, \left ( { G(t) \over G } \, - \, 1 \right ) \, \rho (t) 
\; = \; \third \, \left [ c_\xi \, ( t / \xi )^{1 / \nu} \, + \, \cdots 
\right ] \rho (t) 
\label{eq:p_eff}
\eeq
with $ p_{eff} (t) / \rho_{eff} (t) = \third ( G(t) - G ) / G(t) $.
Strictly speaking, the above results can only be proven if one assumes that
the pressure's time dependence is given by a power law
(as discussed in Section 4).

Equivalently, substituting $ t \approx \alpha R(t) / \dot{R}(t) $, 
one can, as an example, re-write the first Friedman equation as
\beq
{ k \over R^2 (t) } \, + \,
\left ( { \dot{R}(t) \over R(t) } \right )^2 
\; = \; { 8 \pi G \over 3 } \, \left [
1 \, + \, c_\xi \, ( \alpha / \xi )^{1 / \nu} \, 
\left ( { \dot{R}(t) \over R(t) } \right )^{-1 / \nu } \, + \, \cdots 
\right ] \, \rho (t) \, + \, \third \, \Lambda
\label{eq:fried1}
\eeq

The effective Friedman equations of Eqs.~(\ref{eq:fried_tt}) 
and (\ref{eq:fried_rr}) also bear a 
superficial degree of resemblance to what might be obtained
in scalar-tensor theories of gravity \cite{chiba,jordan,dicke} 
(for recent reviews and further references see \cite{cosmrev,bj}),
\beq
S \; = \; \int d x \, \sqrt{g} \, \left [ 
{ 1 \over 16 \pi G} \, f(\phi) \, R \, - \, 
\half \, g^{\mu\nu} \partial_\mu \phi \, \partial_\nu \phi \, - \, V(\phi)
\right ] \; + \; S_{matter}
\eeq
where the 
gravitational Lagrangian is some arbitrary function of the scalar
curvature \cite{inverse}.
It is also well known that often these theories can be re-formulated in terms
of ordinary Einstein gravity coupled appropriately 
to a scalar field \cite{bwhitt}. 
In the FRW case one has for the scalar curvature in terms of the scale factor
\beq
R \; = \; 6 \left( k \, + \, {\dot{R}}^2(t) \, + \, R(t) \ddot{R}(t) \right) / R^2(t)
\eeq
and therefore for $k=0$ and $R(t) = R_0 \, t^{\alpha}$,
\beq
R  \; = \; { 6 \, \alpha ( 2 \, \alpha - 1 ) \over t^2 } 
\eeq
The quantum correction in Eq.~(\ref{eq:fried_tt}) is therefore,
at this level, indistinguishable from an inverse curvature term of the type
$ ( \xi^2 \, R^2 )^{-1 / 2 \nu }$.
But the resemblance is seen here merely as an artifact due to the 
particularly simple form of the RW metric, with the coincidence
of several curvature invariants (see for ex. Eqs.~(\ref{eq:r2})
and (\ref{eq:r22}) ) not expected to be true in general.

Finally let us note that 
the effective field equations incorporating a vacuum-polarization-driven
running of $G$, Eq.~(\ref{eq:naive_t})), 
could potentially run into serious difficulties with experimental
constraints on the time variability of $G$.
These have recently been summarized in \cite{chiba1,gillies,newton,uzan,bies},
where it is
argued on the basis of detailed studies of the cosmic background anisotropy
that the variation of $G$ at the recombination
epoch is constrained as $| G(z=z_{rec}) -G_0 | / G_0 < 0.05 (2 \sigma )$.
Solar system measurements also severely restrict the time variation of
Newton's constant to $ | \dot{G} / G | < 10^{-12} {\rm yr}^{-1}$ \cite{chiba1}.
It would seem though that these constraints can still be accommodated
provided the scale $\xi$ entering the effective field
equations of Eq.~(\ref{eq:naive_t}) is chosen to be sufficiently large,
at least of the order of $\xi > 3 H^{-1} $, given that
in the present model one has
$| \dot{G} /G| \sim {1 \over \nu} \, c_\xi \, t^{1/ \nu -1} / \xi^{1/ \nu}$
and therefore $| \dot{G} /G| \sim 3 \, c_\xi \, t^{2} / \xi^{3}$
for $\nu=1/3$.

\vspace{20pt}

{\bf Acknowledgements}

The authors are grateful to Gabriele Veneziano for his close involvement
in the early stages of this work, and for bringing
to our attention the pioneering work of Vilkovisky and collaborators
on nonlocal effective actions for gauge theories. 
Both authors also wish to thank the Theory Division at CERN
for warm hospitality and generous financial support during the
Summer of 2004.
One of the authors (HWH) wishes to thank James Bjorken for
useful discussions on issues related to the subject
of this paper.
The work of Ruth Williams was supported in part by the UK Particle
Physics and Astronomy Research Council.

\newpage

\vspace{20pt}

\appendix

\section*{Appendix}

\vskip 20pt
\newsection{Classical Field Equations and Conventions}
\hspace*{\parindent}

This appendix is mostly about notation, but also collects a few simple
results used extensively in the rest of the paper.
We will write the Robertson-Walker (RW) metric as
\beq
ds^2 \; = \; - dt^2 + R^2(t) \, \left \{ { dr^2 \over 1 - k\,r^2 } 
+ r^2 \, \left( d\theta^2 + \sin^2 \theta \, d\varphi^2 \right)  \right \}
\eeq
and note that with this choice of signature
(i.e. a minus sign for the $dt^2$ term),
$\Box$ is a positive operator (on functions of t).
Also $\sqrt{g} \equiv \sqrt{-\det(g)} = + r^2 \, \sin \theta \, R^3(t) / \sqrt{1-k r^2}$.

The energy momentum tensor for a perfect fluid is
\beq
T_{\mu\nu} \;  = \; \left( p(t) \, + \, \rho(t) \right) u_\mu \, u_\nu
\, + \, g_{\mu \nu} \, p(t)
\label{eq:perfect}
\eeq
giving in the fluid's rest frame 
$T_{\mu\nu} = diag ( 
\rho, p R^2 / (1-k r^2 ), r^2 p R^2 , r^2 \sin^2 \theta \, p R^2 )$, 
with trace 
\beq
T_\mu^{\; \; \mu} \; = \; 3 \, p(t) - \rho(t)
\eeq
The field equations are then written as
\beq
R_{\mu\nu} - \half \, g_{\mu\nu} \, R \; = \; 8 \pi G \; T_{\mu\nu}
\eeq
The $tt$ component of the Einstein tensor reads
\beq
3 \left( k + {\dot{R}}^2(t) \right) / R^2(t)
\eeq
while the $rr$ component is 
\beq
{ -1 \over 1 - k \, r^2 }
\left( k + {\dot{R}}^2(t) + 2 \, R(t) \, \ddot{R}(t) \right)
\eeq
and the $\theta-\theta$ component
\beq
- r^2 \left( k + {\dot{R}}^2(t) + 2\,R(t)\,\ddot{R}(t) \right)
\eeq
and finally the $\varphi-\varphi$ component
\beq
- r^2 \sin^2 \theta \left( k + {\dot{R}}^2(t) + 2\,R(t)\,\ddot{R}(t) \right)
\eeq
The scalar curvature is simply
\beq
6 \left( k + {\dot{R}}^2(t) + R(t)\,\ddot{R}(t) \right) / R^2(t)
\label{eq:scalar}
\eeq
Thus the $tt$ component of the Einstein equation becomes
\beq
3 \left( k + {\dot{R}}^2(t) \right) / R^2(t) \; = \; 8 \pi G \, \rho (t)
\eeq
while the $rr$ component reads
\beq
{ -1 \over 1 - k \, r^2 }
\left( k + {\dot{R}}^2(t) + 2 \, R(t) \, \ddot{R}(t) \right) \; = \;
8 \pi G \, { 1 \over 1 - k \, r^2 } \; p(t) \, R^2(t)
\eeq
The trace equation is 
\beq
6 \left( k + {\dot{R}}^2(t) + R(t)\,\ddot{R}(t) \right) / R^2(t) \; = \;
8 \pi G \left( \rho (t) - 3 p(t) \right)
\eeq
Covariant conservation of the energy momentum tensor,
$\nabla^{\mu} \, T_{\mu\nu} =0 $ implies
a definite relationship between $R(t)$, $\rho(t)$ and $p(t)$, which reads
\beq
\dot{\rho}(t) + 3 \left( \rho(t) + p(t) \right)
\left( \dot{R}(t) / R(t) \right) \; = \; 0
\eeq
(and which the tensor of Eq.~(\ref{eq:perfect}) in
its most general form does not satisfy).

Next consider the case $k=0$ (spatially flat) and $p=0$ (non-relativistic
matter). 
If $R(t) = R_0 \, t^\alpha$ and $\rho(t) = \rho_0 \, t^\beta$, then
the $tt$ field equation 
\beq
{ 3 \alpha^2 \over t^2 } \; = \; 8 \pi G \, \rho_0 \, t^\beta  
\eeq
implies $\beta = -2$ and $\alpha^2 = 8 \pi G \rho_0 / 3 $,
while the $rr$ field equation
\beq
- \, \alpha \, ( 3 \alpha -2) \, R_0^2 \, t^{2 \alpha -2} \; = \; 0
\eeq
implies $\alpha = 2/3$. 
Also both of these together imply
\beq
\rho(t) \sim t^{-2} \sim ( t^{2/3} )^{-3} \sim 1 / {R(t)}^{3}
\eeq
The trace equation now reads
\beq
{ 6 \alpha \, (2 \alpha - 1) \over t^2 }
\; = \; 8 \pi G \, \rho_0 \, t^\beta  
\eeq
and implies again $\beta = -2$ and $6 \alpha \, (2 \alpha -1) = 8 \pi G \rho_0$.
The latter combined with the $tt$ equation gives 
$3 \alpha^2 = 6 \alpha \, (2 \alpha -1)$, or again $\alpha = 2/3$.
Finally covariant energy conservation implies
\beq
(3 \alpha + \beta ) \, \rho_0 \, t^\beta \; = \; 0
\eeq
or $3 \alpha + \beta = 0$, which does not add to what already comes out of
the $tt$ and $rr$ (or, equivalently, $tt$ and trace) equations, but is
consistent with it.
In conclusion the $tt$ and $rr$ (or $tt$ and trace) equations
are sufficient to determine both $\alpha$ and $\beta$ .

The case of non-vanishing pressure can be dealt with in the same way.
In most instances one is interested in a fairly simple
equation of state $p(t) = \omega \rho(t)$, with $\omega$
a constant. 
For non-relativistic matter $\omega=0$, for radiation $\omega=1/3$,
while the cosmological term can be modeled by $\omega=-1$.
The consistency of the $tt$ and $rr$ equations now requires
\beq
{\alpha \left ( 3 \alpha \, - \, 2 \right ) \over 3 \, \alpha^2 }
\; = \; - \, \omega
\eeq
which gives
\beq
\alpha \; = \; { 2 \over 3 ( 1 \, + \, \omega ) }
\label{eq:alphaomega}
\eeq
for $ - 1 < \omega \leq 1/3 $.
Furthermore from the covariant energy conservation law one has
\beq
3 \, ( 1 \, + \, \omega ) \, \alpha  \, + \, \beta \; = \; 0
\label{eq:betaomega}
\eeq
which implies $\beta = -2 $ again. 
Therefore
\beq
R(t) \; \sim \; t^{ 2 / 3 (1 + \omega ) }
\;\;\;\;\;\;\;\; 
\rho (t) \; \sim \; \left [ R(t) \right ]^{-3 ( 1 + \omega ) }
\eeq
These results are well known and have been collected here
for convenient reference.

\vspace{20pt}

\vskip 20pt
\newsection{Scale Transformations and Gravitational Functional Integral}
\hspace*{\parindent}

Consider the (Euclidean) Einstein-Hilbert action with a cosmological term
\beq
I_E \; = \; \lambda_0 \, \Lambda^4 \int dx \sqrt{g} \, - \, 
\k_0  \, \Lambda^2 \int dx \sqrt{g} \, R
\label{eq:ehaction}
\eeq
Here $\lambda_0$ is the bare cosmological constant,  $\k_0 = 1 / 16 \pi G_0$
with $G_0$ the bare Newton's constant.
Also, and in this section only, we follow customary notation used in
cutoff field theories and
denote by $\Lambda$ an ultraviolet cutoff, not to be confused with 
the scaled cosmological constant.
The natural expectation is for the bare microscopic, dimensionless
couplings to have magnitudes of order one in units of the cutoff, 
$\lambda_0 \sim \k_0 \sim O(1) $.
Next one can rescale the metric
\beq
g_{\mu\nu}' \; = \; \sqrt{\lambda_0} \, g_{\mu\nu}
\;\;\;\;\; 
{g'}^{\mu\nu} \; = \; { 1 \over \sqrt{\lambda_0} } \, g^{\mu\nu}
\eeq
to obtain
\beq
I_E \; = \; \Lambda^4 \int dx \sqrt{g'} \, - \, 
{ \k_0 \over \sqrt{\lambda_0} } \, \Lambda^2 \int dx \sqrt{g'} \, R'
\eeq
Next consider the fact that the (Euclidean) Feynman path integral
\beq
Z \; = \; \int d \mu [g] \; \exp \left \{ 
- \int d x \, \sqrt g \, \Bigl 
( \lambda_0 \, \Lambda^4 \, - \, { \Lambda^2 \over 16 \pi G_0 } \, R
\Bigr ) \right \}
\label{eq:zcont}
\eeq
includes a functional integration
over all metrics, with functional measure ~\cite{dewitt,misner}
\beq
\int d \mu [g] \; = \; \int \prod_x \left ( \det G \right )^{\half} 
\prod_{\mu \geq \nu} d g_{\mu \nu} (x) 
\; = \;  \int \prod_x \; \left ( g(x) \right )^{ (D-4)(D+1)/8 } \;
\prod_{\mu \ge \nu} \, d g_{\mu \nu} (x)
\; \mathrel{\mathop\rightarrow_{ D = 4}} \;
\int \prod_x \prod_{\mu \geq \nu} d g_{\mu \nu} (x) 
\label{eq:dewitt}
\eeq
with the super-metric over metric deformations given by
\beq
G^{\mu \nu, \alpha \beta} (g(x)) \; = \;
\half \; \left ( g(x) \right )^{1/2} \left [
g^{\mu \alpha}(x) g^{\nu \beta}(x) +
g^{\mu \beta}(x) g^{\nu \alpha}(x) + \lambda \,
g^{\mu \nu}(x) g^{\alpha \beta}(x) \right ] \;\; .
\label{eq:dewittsuper}
\eeq
For our purposes it will be sufficient to note that under
a rescaling of the metric 
the functional measure only picks up an irrelevant multiplicative constant.
Such a constant automatically drops out when computing averages.
Equivalently one can view a rescaling of the metric as simply a 
redefinition of the ultraviolet cutoff $\Lambda$,
$\Lambda \rightarrow \lambda_0^{1/4} \Lambda$.
As a consequence, the non-trivial part of the functional
integral over metrics only depends on $\lambda_0$ and
$\k_0$ through the dimensionless combination
$\k_0 / \sqrt{\lambda_0} = 1 / (16 \pi G_0 \sqrt{\lambda_0})$.
The existence of an ultraviolet fixed point is then entirely controlled
by this dimensionless parameter only, both on the lattice ~\cite{critical,ttmodes}
and in the continuum ~\cite{epsilon,litim}.
It is the only relevant (as opposed to marginal or irrelevant,
in statistical mechanics parlance) scaling variable in the pure gravity case, in the
sense of having only one positive (growing) eigenvalue of the linearized renormalization
group transformation in the vicinity of the fixed point.

The parameter $\lambda_0$ controls the overall scale of the
problem (the volume of space-time), while the $\k_0$ term provides the 
necessary derivative or coupling term.
Since the total volume of space-time 
can hardly be considered a physical observable, quantum
averages are computed by dividing out by the total
space-time volume.
For example, for the quantum expectation value of the Ricci scalar one writes
\beq
{ < \int d x \, {\textstyle {\sqrt{g(x)}} \displaystyle} \, R(x) >
\over < \int d x \, {\textstyle {\sqrt{g(x)}} \displaystyle} > }
\label{eq:average}
\eeq
Without any loss of generality one can therefore fix the overall
scale in terms of the ultraviolet cutoff, and
set the bare cosmological constant $\lambda_0$ equal to one in units
of the ultraviolet cutoff. 
\footnote{
These considerations are not dissimilar from the case of a 
self-interacting scalar field where one might want to introduce three
couplings for the kinetic term, the mass term and the quartic
coupling term, respectively. A simple rescaling of the field would then
reveal that only two coupling ratios are in fact relevant.}

The addition of matter field prompts one to do some further rescalings.
Thus for a scalar field with action
\beq
I_S \; = \; \half \, \int dx \sqrt{g} \, \left \{ 
g^{\mu\nu} \, \partial_\mu \, \phi \, \partial_\nu \, \phi \, + \, m_0^2 \, \phi^2
\, + \, R \, \phi^2  \right \}
\eeq
and functional measure
\beq
\int d \mu [\phi] \; = \; \int \prod_x \,
\left ( g(x) \right )^{1/2} \, d \phi (x)
\eeq
the metric rescaling is to be followed by a field rescaling 
\beq
\phi ' (x) \; = \; \phi (x) / \lambda_0^{1/4}
\eeq
with the only surviving change being a rescaling of the bare mass
$m_0 \rightarrow m_0 / \lambda_0^{1/4} $. 
The scalar functional measure acquires an irrelevant multiplicative factor
which does not affect quantum averages.
The bare mass rescaling is of course ineffectual if the fields are massless
to begin with.

The same set of considerations apply as well to the Euclidean
lattice \cite{regge,hartle} regularized version of
Eq.~(\ref{eq:ehaction}), which now reads \cite{hw84,lesh}
\beq 
I_L \; = \;  \lambda_0 \sum_h V_h (l^2) -  2 \, \k_0 \sum_h \delta_h (l^2 ) \, A_h (l^2) 
\label{eq:ilatt} 
\eeq
and
\beq 
Z_L \; = \;  \int d \mu [l^2] \, \; \exp \left \{ 
- \lambda_0 \sum_h V_h (l^2) +  2 \, \k_0 \sum_h \delta_h (l^2 ) \, A_h (l^2)
\right \} 
\label{eq:zlatt} 
\eeq
where, as is customary, the lattice ultraviolet cutoff is set equal to one
(i.e. all lengths and masses are measured in units of the cutoff).
It is known that convergence of the Euclidean lattice functional integral
requires a {\it positive} bare cosmological constant 
$\lambda_0 > 0$ \cite{hw84,lesh,monte}. 

The coupling $\lambda$ should really not be allowed to ``run'', as the overall
space-time volume is intended to be {\it fixed}, not to be itself rescaled under
a renormalization group transformation.
Indeed, in the spirit of Wilson \cite{wilson}, a renormalization group transformation
allows a description of the original physical system in terms
of a new course grained Hamiltonian, whose new operators are
interpreted as describing averages of the original system on the
finest scale, but within the {\it same} physical volume.
This new effective Hamiltonian is still supposed to
describe the original physical system, but does so more
economically in terms of a reduced set of degrees of freedom.
 
The pure gravity theory depends only on one coupling (the dimensionless
$G$), and only that coupling is allowed to run.
This is also, to some extent, implicit in the correct definition of gravitational
averages, for example in Eq.~(\ref{eq:average}).
Physical observable averages such as the one in Eq.~(\ref{eq:average})
in general have some rather non-trivial dependence on the bare coupling
$G_0$, more so in the presence of an ultraviolet fixed point.
Renormalization in the vicinity of the ultraviolet
fixed point invariably leads to the introduction
of a new dynamically generated, non-perturbative scale for $G > G_c$ 
\beq
m \, = \, \xi^{-1} \equiv \, 
\Lambda \, \exp \left ( { - \int^G \, {d G' \over \beta (G') } }
\right )
\, \mathrel{\mathop\sim_{G \rightarrow G_c }} \,
\Lambda \, | \, G - G_c |^{ - 1 / \beta ' (G_c) } \;\;\; .
\label{eq:m_cont}
\eeq
with an exponent related to the derivative of the beta function at the
fixed point 
\beq
\beta ' (G_c) \, = \, - 1/ \nu  \;\; .
\eeq
The overall size of this new scale $\xi$ is controlled by 
the distance from the fixed point $G - G_c $, 
which can be made arbitrarily small
(in the Regge lattice theory one finds for the critical
coupling, in units of the ultraviolet cutoff, $ G_c \approx 0.626$,
and for the exponent $ \nu \approx 0.33 $)

Thus a result such as 
\beq
{ < \int d x \, {\textstyle {\sqrt{g(x)}} \displaystyle} \, R(x) >
\over < \int d x \, {\textstyle {\sqrt{g(x)}} \displaystyle} > } 
\; \sim \; \Lambda^2 \, ( G - G_c )^{\gamma\nu}
\; \sim \; \Lambda^{2 - \gamma} { 1 \over \xi^\gamma }
\label{eq:curvature}
\eeq
referring here to an average curvature on the largest observable scales
(with $\nu$ and $\gamma$ some positive exponents)
does not presumably allow one to state whether the average curvature is large
or small at large distances (that would clearly depend on the choice
of $G-G_c$ and the cutoff $\Lambda$).
\footnote{
Pursuing the analogy with Quantum Chromodynamics, we note
that there the non-perturbative gluon condensate depends in a
nontrivial way on the corresponding confinement scale parameter,
$\alpha_S <F_{\mu\nu} \cdot F^{\mu\nu}> \approx (250 MeV)^4 \sim \xi^{-4} $
with $\xi_{QCD}^{-1} \sim \Lambda_{\overline{MS}}$.} 
But it does establish a definite relationship
between the fundamental scale $\xi$ in Eq.~(\ref{eq:m_cont})
and say the scale of the curvature at the largest scales,
Eq.~(\ref{eq:curvature}), as well as with any other
observable involving $G-G_c$ or $\xi$.
It is the latter curvature that most likely should be identified with
a physical, astrophysically measurable, macroscopic cosmological constant
(and not in any way with $\lambda_0$). 
While it is natural to assume for the curvature measured
on the largest distance scales (for example via the parallel transport
of vectors along very large loops) 
that $R \sim 1/ \xi^2$, and therefore $\gamma=2$, it has proven difficult
so far to establish such a result in the lattice theory, due
to the great technical difficulties involved in measuring small
invariant correlations at large geodesic distances \cite{corr}.

\vspace{20pt}

\vskip 20pt
\newsection{Effective Action Variation}
\hspace*{\parindent}

In this section we will consider the effective gravitational
action of Eq.~(\ref{eq:ieff_sr}), 
\beq
I \; = \; { 1 \over 16 \pi G } \int dx \sqrt{g} \, \sqrt{R} \,
\left( 1 \, - \, A (\Box) \right ) \sqrt{R}
\eeq
and compute its variation.
One needs the following elementary variations
\bea
& & \delta \sqrt{g} \cdot \sqrt{R} \cdot \left( 1 \, - \, A (\Box) \right ) \cdot \sqrt{R}
+ \sqrt{g} \cdot \delta \sqrt{R} \cdot \left( 1 \, - \, A (\Box) \right ) \cdot \sqrt{R}
\nonumber \\
& + & \sqrt{g} \cdot \sqrt{R} \cdot \delta \left( 1 \, - \, A (\Box) \right ) \cdot \sqrt{R}
+ \sqrt{g} \cdot \sqrt{R} \cdot \left( 1 \, - \, A (\Box) \right ) \cdot \delta \sqrt{R}
\eea
Using the identity
\beq
\delta \sqrt{g}  \; = \; - \, \half \, \sqrt{g} \, g_{\mu\nu} \, \delta g^{\mu\nu}
\eeq
as well as $\nabla_{\lambda} \, g_{\mu\nu} =0$ one then has
\bea
& - & \, \half \, \sqrt{g} \, \delta g^{\mu\nu} \, g_{\mu\nu} \, \sqrt{R} \, 
\left( 1 \, - \, A (\Box) \right ) \sqrt{R}
\, + \, \sqrt{g} \, \delta \sqrt{R} \, 
\left( 1 \, - \, A (\Box) \right ) \sqrt{R}
\nonumber \\
& - & \, n \, \sqrt{g} \, \sqrt{R} \, A (\Box) { 1 \over \Box } \, 
(\delta \, \Box) \sqrt{R}
\, + \, \sqrt{g} \, \sqrt{R} \, 
\left( 1 \, - \, A (\Box) \right ) \delta \sqrt{R}
\label{eq:variation}
\eea
Next use is made of the definition of the Ricci scalar,
\beq
\delta R \; = \; g^{\mu\nu} \, \delta R_{\mu\nu} \, + \, 
R_{\mu\nu} \, \delta g^{\mu\nu}
\eeq
For the variation of the affine connection one has
\beq
\delta \Gamma^{\alpha}_{\;\; \mu\nu} \; = \;  \half \, g^{\alpha\beta}
\, \left [ 
\nabla_{\mu} \, \delta g_{\beta\nu} \, + \,
\nabla_{\nu} \, \delta g_{\beta\mu} \, - \,
\nabla_{\beta} \, \delta g_{\mu\nu} \right ]
\label{eq:gammavar}
\eeq
or, equivalently,
\beq
\delta \Gamma^{\alpha}_{\;\; \mu\nu} \; = \; - \, \half \, \left [ 
\nabla_{\mu} ( g_{\nu\lambda} \, \delta g^{\alpha\lambda} ) \, + \,
\nabla_{\nu} ( g_{\mu\lambda} \, \delta g^{\alpha\lambda} ) \, - \,
\nabla_{\beta} ( g_{\mu\kappa} \, g_{\nu\lambda} \, 
g^{\alpha\beta} \, \delta g^{\kappa\lambda} ) \right ]
\eeq
and therefore for the variation of the Ricci tensor
\beq
\delta R_{\mu\nu} \; = \; 
\nabla_{\alpha} \left ( \delta \Gamma^{\alpha}_{\;\; \mu\nu} \right )
\, - \, 
\nabla_{\mu} \left ( \delta \Gamma^{\alpha}_{\;\; \alpha\nu} \right )
\eeq
from which it follows that
\beq
g^{\mu\nu} \, \delta R_{\mu\nu} \; = \; 
\nabla_{\mu} \nabla_{\nu} \left (
\, - \, \delta g^{\mu\nu} \, + \, 
g^{\mu\nu} \, g_{\alpha\beta} \, \delta g^{\alpha\beta} \right )
\; = \; 
g_{\alpha\beta} \, \Box \, \delta g^{\alpha\beta}
\, - \, 
\nabla_{( \mu} \nabla_{\nu )} \, \delta g^{\mu\nu}
\eeq
which is one of the required variations in Eq.~(\ref{eq:variation}).
The second term on the right hand side of the last equation is a
total derivative the ordinary Einstein case, but it needs to be kept here. 
Note also that in general $\Box \, \nabla_\mu \neq \nabla_\mu \Box$, and that
$\Box \, g_{\mu\nu} =0 $ but $\Box \, \delta g_{\mu\nu} \neq 0 $.
For the variation of the covariant d'Alembertian 
\beq
\delta \Box \; = \; \delta g^{\mu\nu} \nabla_\mu \nabla_\nu 
\, - \, g^{\mu\nu} \, \delta \Gamma_{\mu\nu}^{\sigma} \, \nabla_{\sigma}
\eeq
one needs the variation of $\Gamma_{\mu\nu}^{\sigma}$ given by
Eq.~(\ref{eq:gammavar}), which then gives
\beq
\delta \Box \; = \; \delta g^{\mu\nu} \, \nabla_\mu \, \nabla_\nu \, + \,
\nabla_\mu \, \delta g^{\mu\nu} \, \nabla_\nu \, - \,
\half \, \nabla_\mu \, g^{\mu\nu} \, g_{\alpha\beta} \, \delta g^{\alpha\beta} \, \nabla_\nu
\eeq
Here (or at the end) one also needs to properly symmetrize the result for
the variation of $\Box$,
\beq
\delta ( \Box^n ) \, \rightarrow \, \sum_{k=1}^n \, \Box^{k-1}
\, (\delta \Box) \, \Box^{n-k}
\eeq
Next several integrations by parts, involving both the operator
$\Box^n$ (with integer $n$) and well as the operator
$ g_{\mu\nu} \Box - \nabla_{( \mu} \nabla_{\nu )} $,
have to be performed in order to isolate the $\delta g^{\mu\nu}$ term.
This follows from  $\int \sqrt{g} \; \nabla_{\mu} V^{\mu} $ 
$ = \int \sqrt{g} \, (1 / \sqrt{g}) \, \partial_{\mu} \sqrt{g} \, V^{\mu} = 0 $
which allows us to repeatedly integrate by parts and move some
covariant derivatives around.
In general one has to be careful about the ordering of covariant
derivatives, whose commutator is in general non-zero in accord with the Ricci identity
\beq
[ \nabla_{\mu} , \nabla_{\nu} ] \, 
T^{\alpha_1 \, \alpha_2 \dots}_{\;\;\;\;\;\;\;\;\;\;\;\; \beta_1 \, \beta_2 \dots}
\; = \; - \sum_i \, R_{\mu\nu\sigma}^{\;\;\;\;\;\;\; \alpha_i} \,
T^{\alpha_1 \dots \sigma \dots}_{\;\;\;\;\;\;\;\;\;\;\;\;\; \beta_1 \dots}
- \sum_j \, R_{\mu\nu\beta_j}^{\;\;\;\;\;\;\;\;\; \sigma} \,
T^{\alpha_1 \dots}_{\;\;\;\;\;\;\; \beta_1 \dots \sigma \dots}
\eeq
with the $\sigma$ index in $T$ in the $i$-th position in the first
term, and in the $j$-th position in the second term.
The term involving the variation of the covariant d'Alembertian $\Box$ then gives
\beq
- \, n \, \left ( \nabla_\mu \nabla_\nu \sqrt{R} \right ) \,
\left ( { A(\Box) \over \Box } \, \sqrt{R} \right )
\, - \, n \, \left ( \nabla_\mu \sqrt{R} \right ) \,
\left ( \nabla_\nu { A(\Box) \over \Box } \, \sqrt{R} \right )
\, + \, \half \, n \, g_{\mu\nu} \, \left ( \nabla_\alpha \sqrt{R} \right ) \,
g^{\alpha\beta} \, 
\left ( \nabla_\beta { A(\Box) \over \Box } \, \sqrt{R} \right )
\eeq
which again needs to be symmetrized with respect to
$ { A(\Box) \over \Box } \, \sqrt{R} \leftrightarrow \sqrt{R}$, in the
way described above.
After adding the remaining terms, the effective field equations become
\bea
\left ( R_{\mu\nu} \, - \, \half \, g_{\mu\nu} \, R \right ) \,
\left ( 1 \, - \, { 1 \over \sqrt{R} } \, A(\Box) \, \sqrt{R} \right )
\, - \,
\left ( g_{\mu\nu} \, \Box \, - \, \nabla_{( \mu} \nabla_{\nu )} 
\right ) \left ( { 1 \over \sqrt{R} } \, A(\Box) \, \sqrt{R} \right )
\nonumber
\eea
\bea
\, - \, n \, 
\left ( \nabla_\mu \nabla_\nu \sqrt{R} \right ) \,
\left ( { A(\Box) \over \Box } \, \sqrt{R} \right )
\, - \, n \, \left ( \nabla_\mu \sqrt{R} \right ) \,
\left ( \nabla_\nu { A(\Box) \over \Box } \, \sqrt{R} \right )
\nonumber 
\eea
\bea
\, + \, \half \, n \, g_{\mu\nu} \, \left ( \nabla_\alpha \sqrt{R} \right ) \,
g^{\alpha\beta} \, 
\left ( \nabla_\beta { A(\Box) \over \Box } \, \sqrt{R} \right )
\; = \; 8 \pi G \, T_{\mu\nu}
\label{eq:field_var}
\eea
where again the last three terms need to be properly symmetrized
in $ { A(\Box) \over \Box } \, \sqrt{R} \leftrightarrow \sqrt{R}$, as described
above.

Taking the covariant divergence of the l.h.s gives zero for some of the terms,
while the remaining terms give
\bea
- \, \left ( R_{\mu\nu} \, - \, \half \, g_{\mu\nu} \, R \right ) \, 
\nabla^{\mu} \left [ { 1 \over \sqrt{R} } \, \Box^n \, \sqrt{R} \, \right ]
\nonumber
\eea
\bea
\, - \, n \, \nabla^{\mu} \, \left [
\left ( \nabla_\mu \nabla_\nu \sqrt{R} \right ) \,
\left ( \Box^{n-1} \, \sqrt{R} \right )
\, + \, \left ( \nabla_\mu \sqrt{R} \right ) \,
\left ( \nabla_\nu \, \Box^{n-1} \, \sqrt{R} \right )
\, - \, \half \, g_{\mu\nu} \, \left ( \nabla_\alpha \sqrt{R} \right ) \,
g^{\alpha\beta} \, 
\left ( \nabla_\beta \, \Box^{n-1} \, \sqrt{R} \right )
\right ] 
\eea
which has to vanish due to the invariance of the original non-local action.
(Again the last term needs to be symmetrized in
$ \Box^{n-1} \, \sqrt{R} \leftrightarrow \sqrt{R}$).

The above derivation can be slightly generalized to an action of the form
\beq
I \; = \; { 1 \over 16 \pi G } \int dx \sqrt{g} \, R^{1-\alpha}
\left( 1 \, - \, A (\Box) \right ) R^{\alpha}
\eeq
with $\alpha$ a parameter between zero and one (with the previous case
corresponding to $\alpha=1/2$). 
Then for the field equations one obtains an expression of the type
\bea
R_{\mu\nu} \, - \, \half \, g_{\mu\nu} \, R \,
\, + \, \half \, g_{\mu\nu} \, R \, R^{-\alpha} \, A(\Box) \, R^{\alpha}
\, - \, R_{\mu\nu} \, 
\left ( (1-\alpha) \, R^{-\alpha} \, A(\Box) \, R^{\alpha} 
\, + \, \alpha \, R^{\alpha-1} \, A(\Box) \, R^{1-\alpha} \right ) 
\nonumber
\eea
\bea
\, - \, 
\left ( g_{\mu\nu} \, \Box \, - \, \nabla_{( \mu} \nabla_{\nu )} \right ) 
\left ( (1-\alpha) \, R^{-\alpha} \, A(\Box) \, R^{\alpha} 
\, + \, \alpha \, R^{\alpha-1} \, A(\Box) \, R^{1-\alpha} \right )
\nonumber
\eea
\bea
\, - \, n \, 
\left ( \nabla_\mu \nabla_\nu R^{\alpha} \right ) \,
\left ( { A(\Box) \over \Box } \, R^{1-\alpha} \right )
\, - \, n \, \left ( \nabla_\mu \, R^{\alpha} \right ) \,
\left ( \nabla_\nu { A(\Box) \over \Box } \, R^{1-\alpha} \right )
\nonumber
\eea
\bea
\, + \, \half \, n \, g_{\mu\nu} \, \left ( \nabla_\sigma R^{\alpha} \right ) \,
g^{\sigma\rho} \, 
\left ( \nabla_\rho { A(\Box) \over \Box } \, R^{1-\alpha} \right )
\; = \; 8 \pi G \, T_{\mu\nu}
\label{eq:field_var1}
\eea
(where again the last term needs to be symmetrized)
which shows that the choice of either $\alpha=1$ or $\alpha=0$ 
is a bit problematic.

One final question remains, namely what is the relationship between
the above effective field equations, 
Eq.~(\ref{eq:field_var}) or Eq.~(\ref{eq:field_var1}), and the
clearly more economical field equations proposed in Eq.~(\ref{eq:naive_t}).
Obviously the equations obtained above from the variational principle
are much more complicated.
They contain a number of non-trivial terms, some of which are
reminiscent of the $1 \, + \, A (\Box)$ term, and others
with a completely different structure (such as the 
$ g_{\mu\nu} \, \Box \, - \, \nabla_{( \mu} \nabla_{\nu )} $ term).
It is of course possible that when restricted to specific metrics, such
as the RW one, the two sets of equations will ultimately give similar results,
but in general this remains a largely open question.
One possibility is that both sets of effective field equations
describe the same running of the gravitational coupling, 
up to curvature squared (higher derivative) terms, which become irrelevant
at very large distances.

\vfill

\newpage

\end{document}